# Think Fast, Think Slow, Think Critical: Designing an Automated Propaganda Detection Tool

Liudmila, Zavolokina
University of Zurich
Switzerland
zavolokina@ifi.uzh.ch

Kilian, Sprenkamp
University of Zurich
Switzerland
sprenkamp@ifi.uzh.ch

Zoya, Katashinskaya
University of Zurich
Switzerland
zoya.katashinskaya@uzh.ch

Daniel Gordon, Jones
University of Zurich
Switzerland
danielgordon.jones@uzh.ch

Gerhard, Schwabe
University of Zurich
Switzerland
schwabe@ifi.uzh.ch

## ABSTRACT

In today's digital age, characterized by rapid news consumption and increasing vulnerability to propaganda, fostering citizens' critical thinking is crucial for stable democracies. This paper introduces the design of ClarifAI, a novel automated propaganda detection tool designed to nudge readers towards more critical news consumption by activating the analytical mode of thinking, following Kahneman's dual-system theory of cognition. Using Large Language Models, ClarifAI detects propaganda in news articles and provides context-rich explanations, enhancing users' understanding and critical thinking. Our contribution is threefold: first, we propose the design of ClarifAI; second, in an online experiment, we demonstrate that this design effectively encourages news readers to engage in more critical reading; and third, we emphasize the value of explanations for fostering critical thinking. The study thus offers both a practical tool and useful design knowledge for mitigating propaganda in digital news.

## CCS CONCEPTS

• **Human-centered computing** → Human computer interaction (HCI); Empirical studies in HCI.

## KEYWORDS

dual-system thinking, propaganda detection, digital nudging

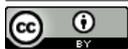

## 1 INTRODUCTION

Effective democracy requires well-informed citizens capable of independent, critical thinking about national and international events. However, many individuals struggle to recognize bias, ideology, slant, and propaganda in the news. As news sources employ increasingly sophisticated logic, many people remain vulnerable to media influences, which largely determine their worldview [1]. Propaganda is the intentional influence of someone's political opinion using various rhetorical and psychological techniques [2]. Propaganda employs methods like loaded language (using words or phrases with strong emotional connotations to influence the audience's opinion) and flag waving (associating oneself or one's cause with patriotism or a national symbol to gain support) [3]. It can be on a spectrum ranging from true to false, harmful or harmless, and appear in various forms, such as online ads and misinformation campaigns. Propaganda, sometimes used for good purposes, often has detrimental effects when employed by political regimes, as seen in events like the Russian aggression in Ukraine [4], [5]. Government-controlled Russian media circulate misinformation, legitimizing Russia's actions in Ukraine, undermining democratic values, and fortifying Putin's regime. The presence of such manipulations and their harm highlight the need for innovative tools for effective propaganda detection and the enhancement of critical thinking.

In today's fast-paced world, news consumption is rapid as well. Many people quickly scan headlines and skim content without diving deep into the substance of the material. Approximately 60% of people favor digital news in text format, mainly for its rapidity [6]. This behavior aligns with and can be explained by Kahneman's dual system theory of cognition, where System 1 is quick and intuitive, and System 2 is slow and analytical [7], [8]. Most of the rapid news consumption happens within System 1, making it susceptible to the influences of biases and propaganda. System 2 is where deeper reflection and critical thinking occur.

Prior research has shown that machine learning (ML) and Natural Language Processing (NLP) methods can be used to detect propaganda or fake news effectively [9], [10], [11]. However, the rise of Large Language Models (LLMs) such as GPT [12], [13], [14], [15], BERT [16], and LLaMA [17] have introduced new capabilities for analyzing diverse content, from news articles to social media posts. Prior research has shown that LLMs can detect propaganda with similar accuracy [18] without resource-intensive training. Furthermore, they can generate context-rich explanations for these detections, thereby deepening users' understanding and fostering their critical thinking by nudging individuals to transition from a surface-level System 1 mode of thinking to a more analytical and critical System 2.

Our research aims to promote a shift from the reflexive System 1 to the analytical System 2, thereby counteracting the negative impacts of propagandistic content on humans' critical thinking. To evaluate our solution, we conducted an expert survey and an online user experiment, examining the solution's potential in real-world contexts.

Rooted in the foundation of Kahneman's dual system thinking [19] and the concept of digital nudges [20], our research adopts the design science research methodology (DSRM) [21] to create a propaganda detection tool, called ClarifAI[1], that fosters critical thinking. DSRM is a widely adopted methodology for developing IT artifacts, ranging from design products and methods to processes. In the context of our project, we developed a proof-of-concept prototype [22] aimed at news readers. Proof-of-concept prototypes serve as practical tools for initial exploration into both the problem space and the usefulness of prospective solutions [22].

The contribution of this study is threefold. Firstly, we present the design of the propaganda detection tool, proposing context-specific nudges and design features to stimulate critical thinking among news readers. These elements are dedicated to triggering behavior change in news readers and pushing them to think more critically. Secondly, in an expert survey and a controlled online user experiment, using the developed Figma prototype and preselected real news articles, we evaluated the tool's usefulness for news readers and its capability to enhance their critical thinking, thereby providing empirical evidence for its real-world applicability. Thirdly, while many research efforts have explored AI-based tools for propaganda detection, our study emphasizes the value of explanations for triggering the intended behavior change, i.e., critical thinking. By revealing the reasons behind the tool's detections, we have enriched the user experience, making the detections more transparent and educational.

The remainder of this paper is organized as follows. Section 2 presents the background literature, covering the three main themes of our research: propaganda detection, dual system thinking, and technology-mediated nudging. The paper then follows the six-step process of DSRM [21], including: (1) problem identification (Section 3), (2) definition of solution objectives (Section 3), (3) design and development (Section 4), (4) demonstration of the prototype (Section 5), (5) its evaluation (Section 6), and (6) communication of the findings, which happens with this conference article. Finally, we discuss the implications of our research and our solution in Section 7 and conclude our research in Section 8.

## 2 BACKGROUND LITERATURE

This study is positioned within three distinct research themes: propaganda detection, dual system thinking, and technology-mediated nudging.

---

[1]The name is tentatively used for this research.

### 2.1 Propaganda detection

Throughout history, propaganda has remained a significant challenge, and with the advent of digital channels, new methods to counteract it have been developed. Initiatives and websites like StopFake.org [23] and investigative organizations, such as Bellingcat [24], FactCheck.org [25], and PolitiFact [26], have been dedicated to fact-checking claims and exposing false information and propaganda to the public. These initiatives, typically involving labor-intensive analysis and research, ensure accuracy, but their manual nature limits scalability.

Though often related, propaganda, misinformation, and fake news are not the same. While misinformation refers to false or misleading information without the intent to harm, and fake news is intentionally fabricated news spread for deceit, propaganda primarily involves manipulating language and using rhetorical techniques to influence public opinion [1]. The focus of this study is on propaganda as a form of language manipulation rather than on fact-checking or verification of news content.

Computational methods can detect propaganda effectively [9], [10], [11]. Persuasion and propaganda detection share many similarities in research [27]. Initially, propaganda detection emphasized document-level analysis, categorizing documents into several categories: trusted, satire, hoax, and propaganda based on their linguistic characteristics [28] and writing styles classifying texts in propaganda vs. non-propaganda [29]. A different research line explored the identification of specific propaganda techniques in texts, primarily in the NLP discipline [2], [3], [27]. Da San Martino et al. [30] created a dataset of news articles annotated with 18 propaganda techniques and explored both technique detection and classification. We use the later version of their propaganda technique classification [31] for our research; the complete list is provided in the Appendix (Table 4). Based on this, the Prta [2] application was developed as a BERT-based tool that evaluates articles based on propaganda techniques with subsequent improvements for more transparent propaganda detection [32]. Though commendable in design, it suffers from frequent false positives and verbosity, impacting user experience. Moreover, as with many advanced algorithms, there is a risk of inherent biases such as political bias [33], [34]. If trained on biased or unrepresentative data, especially in political contexts, these algorithms can unintentionally reinforce and perpetuate such biases during propaganda classification.

Propaganda techniques have expanded their influence beyond traditional media, with recent studies analyzing their prevalence in memes [35] and social media platforms [36]. For a comprehensive review, [3] provide an in-depth survey on computational propaganda detection. Despite the extensive research in the area of propaganda detection, challenges such as the lack of explainability and transparency in automated labeling [37], data scarcity [11], and evaluations often restricted to a single dataset [3] remain and hinder the adoption of such tools. While most current research prioritizes the creation of advanced models for propaganda detection, it often overlooks how end-users receive and perceive these tools. Recent LLMs provide an opportunity for propaganda detection research and practice to enhance user experience and understandability in an often subjective and challenging task of propaganda identification. However, they also come with limitations, such as hallucination issues, often producing inconsistent or unverifiable statements because they rely solely on parametric memory and cannot access external knowledge bases [38]. This overreliance can be problematic when users depend on AI or LLM-generated suggestions for decision-making. Prior research suggests that providing AI-generated explanations can help to reduce this overreliance and increase trust by encouraging users to engage more thoughtfully with the information [39].

There is a notable gap in understanding how propaganda detection tools impact users and change their reading behavior. Research indicates that users' perceptions of automated propaganda detection are varied. Users often trust and concur with fake news detection algorithms, even if they are inaccurate [40]. Yet, users only trust and use these systems if they are transparent and justify their outcome [37]. Research also points to potential unexpected outcomes when tagging fake news articles with warnings [41]. While these warnings reduced trust in false headlines, they unintentionally increased trust in untagged articles, creating an 'implied truth effect' [41]. However, when true headlines were specifically verified, this effect was neutralized. Another study examines the relationship between human assessments and AI predictions in identifying misinformation. As participants engaged with a personalized AI system, trained to understand and adapt to individual user assessments, it was found that while the AI's influence on users' judgments grew over time, it decreased when users rationalized their decisions [42]. Another related yet different stream of research focuses on understanding and improving how people share news. Sharing false news could reinforce its harmful effects by multiplying them. Often, people share false claims without adequately considering their accuracy [43]. To address this, a series of studies [43], [44], [45], [46] have developed and tested various lightweight approaches, specifically focusing on news headlines and users' sharing behaviors. These approaches nudge readers to think about accuracy when sharing news. The approaches include accuracy reminders [44], an accuracy assessment and reasoning for why content is true or false [45] and subtly shifting readers' attention to accuracy is a simple yet effective way to improve choices about what information to share [46]. Altogether, these studies demonstrate that lightweight approaches such as nudges can significantly improve the accuracy of shared news in online environments.

Additionally, the usefulness of human-in-the-loop approaches is shown by several studies [47]. For example, [48] propose that real-time flags from users identifying propaganda can enhance the quality of ML models. While automated propaganda detection holds promise, ensuring users' trust and understanding is essential for its effective adoption. Furthermore, while prior research does not directly measure the influence of automated propaganda detection on critical thinking, it suggests that such methods can help identify and combat propaganda [2], [3], [27], ultimately leading to more informed and critical thinking.

## 2.2 Dual system thinking

Dual process theories describe how we think and decide by categorizing our thought processes into two systems: System 1 (automatic) and System 2 (reflective) [19], [49]. System 1 is intuitive and quick, guiding tasks like driving, while System 2 is more deliberate, taking time to analyze and decide [19]. While System 2 might seem superior due to its thoughtful approach, System 1 also has its benefits. It operates effortlessly, handling multiple tasks simultaneously—like walking, talking, or driving—making it vital for daily routines. The two systems work together, but due to our natural tendency to reduce effort, the reflective System 2 often steps in when System 1 encounters challenges [19]. Nearly 95% of our daily choices are influenced by immediate triggers System 1 produces [20]. However, for important decisions, System 2's critical thinking becomes crucial. Yet, its limitation is that it can easily get overwhelmed due to its intensive cognitive demands.

When faced with multiple demanding tasks, our brain might switch to the more automatic System 1, even for decisions better suited for System 2. This shift can lead us to rely on shortcuts or heuristics, potentially overlooking critical judgment errors [20]. These shortcuts can sometimes lead to biases. For example, status-quo bias refers to a preference for existing conditions over new ones [50] and confirmation bias refers to our inclination to primarily seek information that aligns with our existing beliefs [20]. The original studies of Kahneman [7], [8], [19] describe six distinct characteristics of dual system thinking, we summarize them in Table 1. In our study, these characteristics inform the understanding and definition of critical thinking, i.e., the degree to which System 1 or System 2 is active.

Despite the widespread acceptance of Kahneman's dual system thinking theory and extensive empirical evidence across various fields [51], it has been criticized in several ways [51]. One major criticism relates to oversimplifying human cognition by breaking it down into two distinct systems. Researchers argue that the interplay of cognitive processes is more complex and goes beyond Kahneman's binary classification with a clear-cut distinction between the two systems. Instead, there is a much stronger overlap between their functions. Moreover, the characteristics typically associated with the two systems are not consistently observed together or are suggested to be seen as a continuum instead of discrete binary characteristics, challenging the existence of two distinct cognitive systems [51]. Another criticism relates to labeling biases as errors or fallacies in Kahneman and Tversky's work [52]. This criticism suggests that intuitive processes should not be considered incorrect per se because the norms used to judge are complex and not always applicable in every scenario [52].

In the field of online news consumption, prior research has designed tools to prime users to rely less on intuition and think more critically about content [43], [44], [45], [46], [53]. One study [54] explored the implications of dual system theory, particularly in differentiating between real and fake news. Their results indicated a significant relationship between analytical reasoning skills and the ability to accurately identify fake news [54]. This study calls for new intervention designs to tackle misinformation and the spread of fake news, showing the usefulness of dual system theory extending beyond its traditional areas of application in behavioral economics or psychology.

## 2.3 Technology-mediated nudging

Propaganda often exploits our automatic System 1 response, making us prone to its influence. To counteract this, it is crucial to encourage users to engage their more reflective System 2 thinking when they encounter potential propaganda. However, propaganda goes beyond just providing information. It influences psychological processes. Hence, there is a need for mechanisms that address these underlying processes. To trigger behavior change, 'nudges' can be embedded in the design [20]. The nudge is a concept stemming from behavioral economics. Nudges are subtle cues that can alter human behavior based on knowledge of cognitive biases to improve humans' decisions without restricting them or drastically changing incentives [20], [55]. The use of nudges has expanded across fields, including Human-Computer Interaction (HCI).

Table 1: Dual system thinking characteristics (based on [7], [8], [19])

| Characteristic | Definition | System 1 | System 2 |
| --- | --- | --- | --- |
| Speed | Speed of thinking | Fast | Slow |
| Processing | Approach to handling thinking tasks | Parallel | Serial |
| Control | Degree of conscious oversight | Automatic | Controlled |
| Effort | Cognitive load | Effortless | Effortful |
| Nature | Inherent operating mechanism | Associative | Rule-governed |
| Adaptability | Ability to change or evolve | Slow-learning | Flexible |

In HCI, designers of systems implement digital nudges that aim to alter user behavior to improve user experience or choices. Digital nudges can target either the automatic or reflective processes and be transparent or non-transparent to the user [20], [56]. Prior research collected a useful set of 23 digital nudges, classified them into three categories, facilitator, spark, and signal, and described how they could be used in system design for different purposes [20], [57]. Their classification is based on Fogg's behavior model for persuasive design, which suggests that an individual needs enough motivation, ability, and a timely trigger for a desired behavior to occur [57]. *Facilitator* nudges simplify tasks, aiding motivated users who find execution challenging (i.e., lack the ability) [20]. For instance, setting default options can reduce cognitive effort, while other facilitator nudges encourage reflective choices by suggesting alternatives. *Spark* nudges aim to increase motivation in users who have the ability but lack the motivation. They increase motivation by emphasizing potential losses, supporting public commitments, offering competitive alternatives, or reinforcing social acceptance mechanisms. *Signal* nudges address situations where users' actions differ from their intentions, even when both motivation and ability are in place. These nudges introduce doubt, create discomfort with the ongoing action, or attract attraction to certain stimuli, guiding users towards desired behaviors [20].

Digital nudges, especially when applied subtly, may effectively mitigate propaganda and misinformation [55], [58], [59]. However, so far, the existing empirical evidence on the effectiveness of nudging in the context of news consumption remains limited. This topic requires further research, particularly regarding the practical application of digital solutions and tools that incorporate nudging techniques. Instead of solely seeking facts or discrediting unreliable news sources, digital solutions should nudge users towards more informed and critical news consumption [53]. Over time, this repeated exposure can foster trust in more credible information sources and ultimately improve societal outcomes [53]. This approach, supported by prior research [43], [44], [45], [46], demonstrates that lightweight nudges can significantly improve the accuracy of news distribution in online environments. Furthermore, integrating nudges in digital tools available to users can enhance user engagement and accuracy in combating misinformation [60], [61]. To effectively implement these nudges, employing well-established web design techniques, such as browser extensions that offer text highlights and alerts, can be particularly useful [62], [63], [64] and change user reading behavior online.

An analysis of 100 empirical studies [65] shows that 62% of nudging treatments yield statistically significant results. Nudges typically have a median effect size of 21%, with this impact varying based on the type of nudge and application context [65]. Among different types of nudges, defaults are the most effective type of nudge, whereas precommitment strategies are the least [65]. As for application context, nudges are most effective in privacy, energy, and finance [65]. However, the literature review did not examine the context of news consumption. This omission likely stems from the fact that the studies analyzed did not include this particular application area. Furthermore, digital nudging, while similarly effective as traditional nudging, provides new opportunities for individualized approaches [65].

However, nudges may also fail for various reasons [20]. They may lack lasting educational impact, with their effects being temporary and not persisting over time [20]. Some nudges may also cause unintended behaviors or user mistrust. Others, perceived as intrusive, can prompt users to resist or rebel against the intended action, especially when their timing or intensity is misjudged [20]. Moreover, nudges may also be ineffective when faced with individuals' strong preferences or deeply ingrained habits that nudges cannot change [20]. Despite their potential, these challenges can inhibit the effectiveness of nudges, indicating the need for a more context-specific approach [20].

## 3 PROBLEM IDENTIFICATION AND OBJECTIVES FOR THE SOLUTION

The first step in the DSRM is to "*define the specific research problem and justify the value of a solution*," followed by the second step of describing solution objectives based on the problem definition and our understanding of feasible possibilities [21]. We identified two main problems, which we describe below. Subsequently, based on dual system thinking, we outline two *design goals* for a solution (an artifact) addressing the mentioned problems.

## 3.1 Difficulty in identifying propaganda

*Difficulty in identifying propaganda (P1)* is the first problem to be addressed. Propaganda aims to influence opinions, and it often blends into regular content. While some propaganda is direct and easy to identify, others are more subtle and can be harder to spot [66]. The strategies used to spread such messages and the readers' biases often converge, complicating the task of separating truth from manipulation [67]. Even after readers are presented with accurate information, they might still remember and be influenced by the initial incorrect details, known as the 'continued influence effect' [68], [69]. In the presence of large amounts of information, users are often overwhelmed, making rapid judgments without delving deep into the content, which can result in unconsciously consuming propaganda [7], [19]. This leads us to the definition of our first *design goal*:

*DG1. Intuitive interaction (System 1): Ensure users can quickly identify propaganda with ease.* To align with System 1's rapid processing, our goal is to develop a tool to facilitate the immediate detection of propaganda. This way, even during a quick scroll, users should be able to spot potential propaganda cues. To achieve this, the design should be non-intrusive, ensuring that while the cues are visible, they do not hinder the user's natural reading flow. So, a balance between being noticeable yet not disruptive is essential.

## 3.2 Rapid and uncritical news consumption

*Rapid and uncritical news consumption (P2)* is the second problem to be addressed. The current digital consumption patterns largely engage System 1, where quick and impression-based decision-making is dominant. When reading news, System 1 quickly forms impressions from headlines or visuals, making them influence and shape opinions, especially when messages are repeated, enhancing their perceived truthfulness [70]. Emotional content, often used in propaganda, engages System 1, making messages memorable [71]. Meanwhile, System 2 evaluates news critically but can be lazy [7], [19]. When it aligns with existing beliefs, there is a tendency not to challenge the information, which leads to echo chambers on platforms like Facebook [72]. Effective propaganda relies on System 1's reactivity while bypassing System 2's criticality. This processing mode becomes particularly concerning when considering the proliferation of misleading and biased news. The rapid and uncritical consumption of information means that many users can, and often do, internalize propaganda without any conscious realization of doing so. While spotting propaganda is the first step, comprehending it is equally important. Thus, the second design goal is to encourage users to shift from a passive consumption mode into an analytical and reflective one. Therefore, we formulate the second *design goal* as follows:

*DG2. Critical analysis (System 2): Facilitate a deeper understanding and reflection during news consumption.* This means that once a potential piece of propaganda is identified, the user should have the means to delve deeper, understanding the nuances and techniques of the propaganda. The tools and features we design should not just raise an alarm but also make transparent to the user the 'why' and 'how' behind the detected propaganda, promoting more informed news consumption.

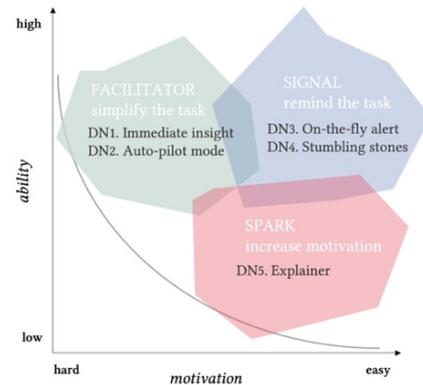

**Figure 1: Digital nudges for propaganda detection on Fogg's behavior model for persuasive design**

## 4 DESIGN AND DEVELOPMENT

After defining our design goals for a solution, we proceed to the third step of the DSRM [21]: designing and developing the artifact, i.e., our propaganda detection tool, ClarifAI. In a nutshell, ClarifAI provides real-time analysis of news content, prompting users to perceive the information they consume critically. ClarifAI detects and highlights instances of propaganda in a text, alerting the user about the potential presence of propaganda and explaining why it is considered as such. In this section, we describe the design and developed features in detail.

We developed a set of digital nudges to alter the reading behavior. Table 2 summarizes the problems (P), design goals (DG), digital nudges (DN), and the implemented features (IF). The DNs were derived based on the DGs and the nudging techniques [20] for effective persuasion to make sure we address all three aspects of Fogg's Behavior Model. The IFs were then defined based on well-established approaches for similar tools for web content, such as clickbait or fake news identification [62], [63], [64]. We aligned the proposed DNs according to Fogg's Behavior Model for persuasive design, classifying them as facilitator, signal, or spark nudges. This alignment is illustrated in Figure 1. The high-level architecture, the process of how the ClarifAI extension works, and the corresponding implemented features (IFs) are shown in Figure 2, while the user interaction is demonstrated in Figure 3 and Figure 4.

### 4.1 Immediate insight

To simplify propaganda detection and help users who are motivated to recognize propaganda in the news but might find it challenging, we propose *DN1 (Immediate insight)* as a facilitator nudge [20]. This nudge embeds the propaganda detection tool within the user's natural browsing environment, ensuring immediate and effortless access to propaganda detection capabilities. As for the implementation, two options were considered: an online web application and a browser extension. The idea of a web application, i.e., a website where a user could copy-paste a piece of news to be analyzed and then view propaganda detected in the text, was disregarded. This was because users, while reading quickly, probably would not want to go to an external website due to the additional effort required. So, to ensure that the use of ClarifAI is as easy as possible and embedded in the natural reading environment (e.g., news portals), we opted for designing ClarifAI as a browser extension. This DN was implemented through *IF1 (Browser extension)*.

Table 2: Problems, design goals, digital nudges, and implemented features

| Problem | Design goals | Digital nudges | Implemented features |
| --- | --- | --- | --- |
| **P1: Difficulty in identifying propaganda** | **DG1. Intuitive interaction (System 1)**: Ensure users can quickly identify propaganda with ease. | **DN1. Immediate insight**: This nudge embeds the propaganda detection tool within the user's natural browsing environment, ensuring immediate and effortless access to propaganda detection capabilities. | **IF1. Browser extension**: Automatically analyzes text content without the need for external applications. |
|  |  | **DN2. Auto-pilot mode**: By making the tool's default behavior to highlight potential propaganda, users are nudged into using the propaganda detection tool without any additional effort. | **IF2. 'On' as a default setting:** Ensures automatic and continuous propaganda detection, integrating the use into users' routines without manual activation. |
|  |  | **DN3. On-the-fly alert**: By providing detection results during active browsing, this nudge maintains the user's current context and facilitates rapid cognitive reaction when potential propaganda is detected. | **IF3. Real-time LLM-based detection:** Offering real-time LLM-based detection as users browse, reinforcing the transition from System 1 to System 2 thinking. |
| **P2: Rapid and uncritical news consumption** | **DG2. Critical analysis (System 2)**: Facilitate a deeper understanding and reflection during news consumption. | **DN4. Stumbling stones**: Highlights of propaganda with distinctive colors help users easily identify content that may require a more critical review, acting as a 'stumbling stone' in their reading flow. | **IF4. Content flags:** Flag propaganda content with colors that stand out to make detections noticeable, thereby emphasizing the presence of propaganda in the text. |
|  |  | **DN5. Explainer**: By offering an explanation during the interaction, this nudge facilitates an understanding of a detected propaganda technique, helping to consciously recognize manipulation tactics. | **IF5. LLM-generated explanations**: When a user hovers over a flagged segment, the extension displays LLM-generated clarification, explaining the propaganda technique used and why the content may be propagandistic. |

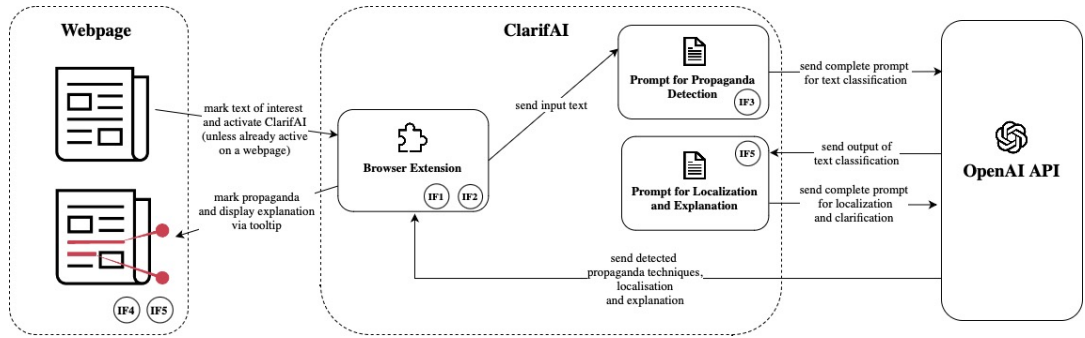

Figure 2: Architecture of ClarifAI

*IF1. Browser extension.* ClarifAI is designed to operate as a browser extension, automatically analyzing text content without the need for external applications. The extension is thus easily distributable across different hardware and web browsers. This way, the entrance barrier for users is low, making it easy to install.

## 4.2 Auto-pilot mode

To remove any barriers to using this propaganda detection tool, we propose *DN2 (Auto-pilot mode)* as a facilitator nudge [20]. By making the tool's default behavior to detect potential propaganda, users are nudged into using it without any additional effort. This automatic detection ensures that users do not have to manually switch on the tool when browsing their regular news sources. With this nudge, we wanted to circumvent the potential forgetfulness or laziness of the user and minimize disruption to the natural reading environment, so the auto-pilot mode works in the background. By default, the automatic detection is activated. However, if the automatic detection is turned off, users can still select the specific text they want to analyze. Upon text selection, the tool automatically flags potential propaganda without further action from the user. This DN was implemented through *IF2 ('On' as a default setting)*.

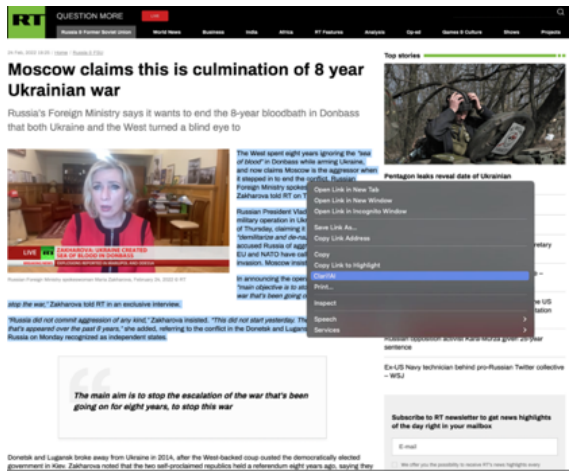

Figure 3: Interaction with ClarifAI: the browser extension (IF1) with text selection (in case the automatic detection, IF2, is turned off)

*IF2. 'On' as a default setting.* Ensures automatic and continuous propaganda detection, integrating the use into users' routines without manual activation. After the user installs the extension, the ClarifAI extension will be shown within the browser and is always available. The user can then activate the propaganda detection either on the whole webpage, whereby the extension scans the relevant HTML attributes, or mark the text themselves, which ClarifAI then processes.

### 4.3 On-the-fly alert

To address situations where users' actions may differ from their intentions when reading news articles, even when both motivation and ability are in place, we implemented *DN3 (On-the-fly alert)* as a signal nudge [20]. This nudge provides propaganda detection results during active browsing, maintaining the user's current context and facilitating rapid cognitive reaction when potential propaganda is detected. It serves as a trigger, guiding users toward a more critical evaluation of content. Thus, this signal nudge ensures a more reflective and cautious engagement with news, enhancing the overall reading experience. This DN was implemented through *IF3 (Real-time LLM-based detection).IF3. Real-time LLM-based detection.* The LLM-propaganda detection system operates through a structured two-step mechanism, real-time detection, and LLM-generated explanations (see Figure 2). Within each step, a dedicated prompt has been designed, which interacts with the API of GPT-4 [15]. For real-time detection, the system leverages few-shot prompt-based learning [14], meaning we provide the LLM with an explanation as well as an example of each of the given propaganda techniques (Table 4 in the Appendix). Further, we employ the chain of thought prompt technique, meaning that we let the model reason about each occurrence of a propaganda technique. Prior research found that we can steer the model into a more conservative direction, only predicting an instance of propaganda if the model can reason about it [18]. Within *IF3*, we solely use the reasoning to steer the model output, while the explanation shown by the tool is generated in *IF5*. We then obtain a list of propaganda techniques detected in the given article from the LLM. The prompt for the detection of propaganda can be found in Table 5 in the Appendix.

### 4.4 Stumbling stones

To further support users, we have implemented *DN4 (Stumbling stones)* as a signal nudge [20]. This nudge adds highlights to potential propaganda with distinctive colors, acting as a 'stumbling stone' in their reading flow. The color highlights prompt users to look closely at specific content flagged as propaganda. Unlike an intrusive popup or other disruptive alert, this method allows users to continue reading at their own pace but with increased awareness. Thus, the stumbling stones serve as subtle yet effective reminders, urging users to approach certain information cautiously. This DN was implemented through *IF4 (Content flags).IF4. Content flags.* To make propaganda detections noticeable and thereby emphasize the presence of propaganda in the text, ClarifAI flags propaganda content with colors. As part of the browser extension, this feature assigns specific color codes to the corresponding text, creating a visual highlight within the webpage content once propaganda is detected. These colors are chosen to stand out from the regular text and background, attracting the reader's attention without being overly disruptive.

### 4.5 Explainer

Finally, we propose *DN5 (Explainer)* as a spark nudge [20]. This nudge offers explanations during the interaction, describing in more detail the detected propaganda techniques and why they are considered as such. The aim is to foster a deeper understanding and critical thinking, especially for those users who may have the ability to discern but lack the motivation to engage with the content critically. By providing context-specific explanations within the natural browsing environment, users can grasp the essence of the propaganda technique used without the need to navigate away from their reading flow. This approach makes propaganda detection educational and motivates users to be more proactive in understanding the news they consume. This DN was implemented through *IF5 (LLM-generated explanations)*.

*IF5. LLM-generated explanations.* When a user hovers over a flagged segment, the extension displays LLM-generated clarification, explaining the propaganda technique used and why the content may be propagandistic. Similarly to *IF3 (Real-time detection)*, we employ a prompt (see Appendix, Table 5) to locate the previously identified propaganda technique within the article. Further,

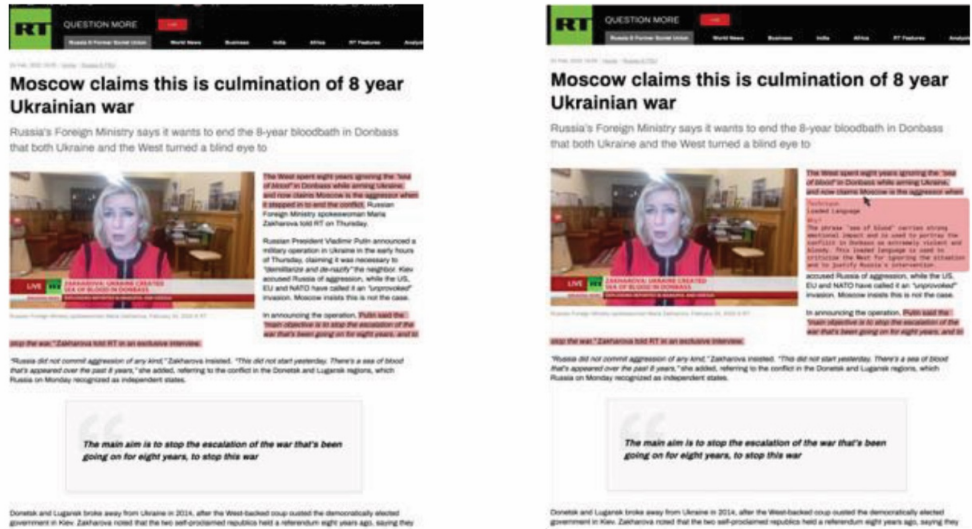

Figure 4: Interaction with ClarifAI: real-time LLM-based propaganda detection (IF3) with content flags (IF4) (left) and LLM-generated explanations (IF5) (right)

we prompt the LLM to clarify why the given text section is propagandistic. The LLM then generates an explanation for the previously identified section in relation to the complete article. The obtained information on the propaganda technique, location in the article, and explanation are fed back to the browser extension.

## 5 DEMONSTRATION

The fourth step in the DSRM is the demonstration of the proposed artifact [21]. We do so by 1) gathering expert feedback on the quality of explanations in an online survey to demonstrate and evaluate that such design is possible and that LLMs produce useful explanations, and 2) conducting an online experiment where potential users can experience ClarifAI in a controlled experimental setting to evaluate the attainment of our design goals, DG1 and DG2. This study received ethical approval from an Institutional Review Board (IRB) under the ID OEC IRB #2023-076. In this section, we describe the participants of the expert survey and the online experiment, the procedure, and the data collection.

### 5.1 Participants

**Expert survey.** For the expert survey, the participants were recruited through an online platform, Prolific [73], which is frequently used in academic research and is known for its reliable participant base [74]. The participants were offered a monetary incentive of 18.75 GBP. For the recruitment, we set several inclusion criteria: the participants 1) have professional media or political studies background with at least an M.Sc. or Ph.D. degree, which ensures that we capture professional perceptions; 2) are native English speakers. Expert 1 is a 23-year-old female based in the United Kingdom with a graduate degree in Politics, Law, and Finance. Expert 2 is a 25-year-old male also based in the United Kingdom, holding a graduate degree in Politics. The average completion time was 4 hours and 7.5 minutes.

**Online experiment.** We again used Prolific [73] to recruit participants for the online experiment. The participants were offered a monetary incentive of 9 GBP/hour to participate in the study. They were a diverse group with varying levels of gender, age, education, and country of residence (see the demographics in Appendix, Table 6). We also used the Need for Cognition Scale [75], a psychological tool used to measure an individual's propensity for engaging in and enjoying thinking, to assess the overall tendency of the participants' thinking. During the recruitment, we set several exclusion criteria: 1) the participants have no professional media or political studies background, which ensures that we capture general perceptions rather than professional ones; 2) only native English speakers were selected to participate in the study to ensure the participants have comparable levels of English proficiency; and 3) the participants could only use desktop devices. The experiment was pretested with 11 participants. The experiment was completed by 248 participants out of 327 who started it, making the dropout rate 24%. The participants could complete the experiment from the location of their choice since the experiment was conducted entirely online.

### 5.2 Procedure

**Expert survey.** The experts received instructions for their task, i.e., to evaluate the quality of propaganda detection and explanations, a briefing about 14 propaganda techniques (Table 4), and the assessment sheet in an Excel file. The assessment sheet included 20 articles focused on a current topic, the ongoing war in Ukraine, sourced from outlets known for propagandistic content [76], [77], namely Russia Today [78] and Infowars [79]. In total, ClarifAI flagged 107 instances of propaganda in 20 articles. These instances were subsequently reviewed by experts using an assessment sheet. On this sheet, experts were given the article text, a passage marked as propaganda with the identified propaganda technique, and its explanation generated by ClarifAI. For each, the experts recorded their evaluations of both the detection accuracy and the explanation quality. Upon completion, they submitted the filled-out assessment sheet back to the researchers.

**Online experiment.** The experiment was conducted following a between-subject design, meaning each participant was assigned to a single treatment. Once participants had provided informed consent, they were asked about their demographic data and randomly assigned to one of the three groups:
(1) no propaganda detection tool (Basic),
(2) propaganda detection tool without explanations (Light), and
(3) propaganda detection tool with explanations (Full).

Each participant then read two news articles that contained elements of propaganda. Depending on the group (2 & 3), they used the propaganda detection tool that either only marks detected propaganda (Light) or also provides explanations of propaganda (Full). The control group (Basic) received unmarked texts. After reading each of the articles, participants completed an online survey. All three groups were asked to answer the same survey questions that related to the entire texts of the articles. The reading task was formulated as follows, with the text marked in *italics* only seen by groups Light and Full since the tool was not present for group Basic.

> "Further, you will be asked to read two news articles. Each article may contain elements of propaganda. *You will also interact with a propaganda detection tool as you read. The propaganda detection tool may identify and mark some manipulation techniques. Please engage with these in a way that feels natural to you.* After reading, you will answer a series of questions related to your experience and your impressions of the news content."

To ensure the quality of the answers provided by the participants, a commitment request and screener validation questions were integrated into the flow of the experiment [80], [81]. After completing the reading task and answering the survey questions, the participants received a debriefing, where they could learn more about the details of the experiment.

For the experiment, we chose three real news articles (see Appendix, Table 7). These articles were sourced again from Russia Today [78] and Infowars [79] and were also in the articles assessed by experts described above. Each participant was randomly assigned two out of these three articles to read. To control for readability and ensure no effects were due to differences in comprehension level, we verified that all three articles had the same Flesch-Kincaid Grade Level [82] of 12. ClarifAI showed the detection of three propagandistic passages in each article and explanations for them. The entire experiment was set up in Qualtrics [83] reducing technical setup effort for the participants as they did not have to switch between different applications. To ensure ease of use during the experiment and eliminate the need to install the extension, we mocked the functionality of ClarifAI with Figma prototypes for all three participant groups. However, we did not manipulate the real articles. Furthermore, all the detections and clarifications were generated by the functional version of ClarifAI, which we then transferred to the Figma prototypes.

For each of the articles, three prototypes were created: Basic (without propaganda detections), Light (only detections are present without explanations), and Full (both detections and explanations are present). Figure 5 shows how the prototypes looked like for each of the groups, also showing the three selected articles. To access all nine prototypes (Basic, Light, and Full per article), see Appendix, Table 7. The Figma prototypes were embedded in the survey via iFrame, an HTML element that allows embedding another HTML document within a webpage. This allowed participants to see a news webpage directly within the survey page. The average completion time of the experiment was 12 minutes (Basic), 13 minutes (Light), and 21 minutes (Full).

### 5.3 Data collection and analysis

**Expert survey.** The primary source of data collection from the experts was the survey questions, defined based on media studies literature [41], to assess the quality of the explanations. The expert survey evaluated whether the generated explanations were accurate and informative, ensuring they were not randomly fabricated but useful for understanding propaganda techniques. However, we did not control for false negatives (i.e., whether the ClarifAI missed some propaganda instances) since we consider the impact of false negatives less significant than that of false positives in the context of the study. We used the following criteria:

- **Correctness of propaganda technique detection:** The experts were asked to assess ClarifAI's ability to correctly identify propaganda techniques, choosing between 'Yes' or 'No' indicating whether they agree with the identification. If they disagreed, they were asked to specify what they thought was the correct technique or whether it was not propagandistic at all.
- **Explanation quality**: To assess the quality of the explanations, we used such criteria as accuracy, informativeness, believability, and clarity. For all items, we used a 7-point Likert scale, ranging from 1 – not at all to 7 – very much.

**Online experiment.** The primary source of data collection here was the survey questions the participants were asked to respond to after they had read the articles and interacted with the different versions of ClarifAI. The survey consisted of multiple sections to capture various dimensions of participants' perceptions and experiences and evaluate whether we attained our design goals DG1 and DG2. To ensure that both design goals can be effectively evaluated, we specifically focused on five metrics. Table 3 summarizes how each design goal relates to the selected evaluation metrics.

For DG1, i.e., 'ensure users can quickly identify propaganda with ease,' two metrics were selected:

- **Propaganda Awareness:** The immediate awareness of the presence of propaganda is central to our study. The participants using the 'Light' and 'Full' versions of ClarifAI were asked a Yes/No question about whether the tool increased their awareness of propaganda techniques.
- **Net Promoter Score (NPS)**: NPS is widely used to reflect customer, or user, experience and satisfaction [84]. NPS was

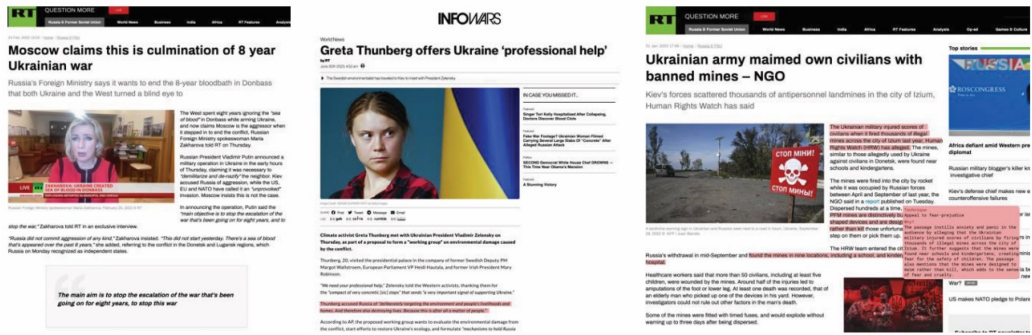

Figure 5: Examples of the prototypes for groups Basic (left), Light (center), and Full (right)

used to assess if the tool successfully makes users engage more critically to the extent that they would recommend it to others. The participants using the 'Light' and 'Full' versions of ClarifAI were asked to rate their likelihood of recommending the ClarifAI tool on an NPS scale from 0 to 10. Based on this question, participants can be described as detractors (score 0-6), passives (score 7-8), or promoters (score 9-10). The score is calculated by subtracting the percentage of Detractors from the percentage of Promoters. Passives are not included in the score but play a role in the overall distribution.

For DG2, i.e., 'Facilitate a deeper understanding and reflection during news consumption,' four metrics were selected:

- **Slow vs. Fast thinking:** Based on Daniel Kahneman's model of dual-system thinking [8] and the specific characteristics in Table 1, we formulated questions to capture participants' thought processes while reading articles. The scale ranged from 1 to 7, serving as a proxy for more 'intuitive' or 'analytical' approaches to consuming news. To ensure the internal validity, we calculated Cronbach's Alpha value, $\alpha = 0.66$, indicating acceptable reliability, for six items. The overall value of the thinking mode variable was calculated as mean across the six variables. We assess the individual impacts of each of the six constituting variables, as well as the overall value, to understand their independent contributions to the overall construct. This approach allows us to analyze each variable separately rather than interactions between variables.
- **Reading time** spent per article: This offered insights into whether the presence of the propaganda detection tool influenced reading speed and engagement with the articles. We consider reading time an appropriate measure reflecting slower and more deliberate reading.
- **Perception of bias in the news:** This question captures how users feel about the objectivity of the news article. Again, we used a 7-point Likert scale, ranging from 1 – not at all to 7 – very much.

- **Net Promoter Score (NPS)**, as described above since it is relevant for both design goals as a metric of the overall user satisfaction.

For data analysis, we conducted a one-way Analysis of Variance (ANOVA) to examine the differences across the groups: Basic, Light, and Full. ANOVA was chosen over multivariate analysis of variance (MANOVA), which would allow for examining interactions among multiple dependent variables, because our study's design does not necessitate this level of complexity. Instead, ANOVA treats each variable independently, which aligns with examining distinct cognitive characteristics separately, as is often done in dual system theory [51]. This decision was also supported for the sake of clarity and interpretability of results. Post-hoc t-tests were conducted to identify where these differences lay. The analysis was performed with the Python statistics package SciPy. Finally, we allowed the participants to leave their comments in a free-text field, 'Additional comments' in the survey. Altogether, we received 51 free-text comments that provided further insight and ideas on using ClarifAI. These were grouped according to their themes for each experiment group. Finally, we estimate the potential costs of using ClarifAI to give insight into practical feasibility of such implementation.

## 6 EVALUATION

The fifth step in the DSRM involves assessing how well the proposed solution and its design address the identified problems [21]. For instance, the evaluation might compare the artifact's functionality against the intended solution objectives. In our case, we first report the results of the online experiment and then the expert survey.

### 6.1 Propaganda awareness and Net Promoter Score

**Propaganda awareness:** The results of the propaganda awareness question show that both the Light and Full versions of ClarifAI are good at making the participants more aware of propaganda techniques. But the Full version does an even better job. In the Light version, 82% of the participants said they became more aware of propaganda, while in the Full version, 96% said it helped them become more aware of propaganda techniques. The results are visualized in Figure 6. The Full group showed significantly higher propaganda awareness with M = 0.96 compared to the Light group with M = 0.82, with the ANOVA indicating a statistically significant difference with a p-value of 0.003 (p < 0.01). The detailed statistical results can be found in Appendix (Table 10).

Table 3: Design goals and the metrics for the goal attainment (★ indicates a relationship)

| Design Goals \ Metrics | Critical Thinking | | Perception of bias | Propaganda awareness | Satisfaction (Net Promoter Score) |
|---|---|---|---|---|---|
| | Slow vs. Fast thinking | Reading time | | | |
| DG1. Intuitive interaction (System 1): Ensure users can quickly identify propaganda with ease. | | | | ★ | ★ |
| DG2. Critical analysis (System 2): Facilitate a deeper understanding and reflection during news consumption. | ★ | ★ | ★ | | ★ |

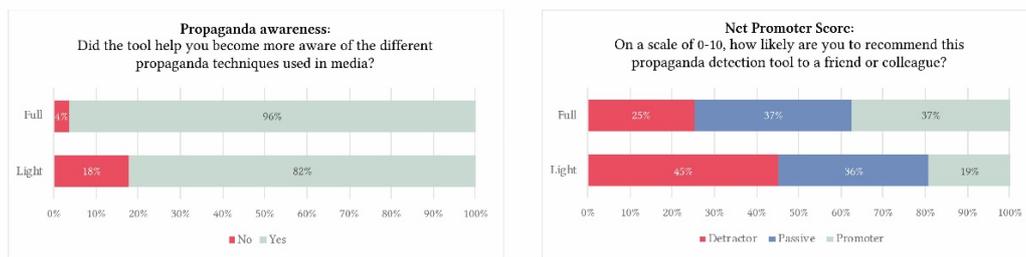

Figure 6: Propaganda awareness (left) and Net Promoter score (right)

**Net Promoter Score:** The NPS score for the Light version is -26, with a predominance of Detractors. The Full version scored better with an NPS score of 12 with a more balanced percentage of Passives and Promoters, and fewer Detractors. The Full version, including explanations and propaganda detection, received more favorable opinions, suggesting that the added explanations (DN5) contributed to a better user experience. The results are visualized in Figure 6. The Full group also scored higher on NPS with M = 7.49 than the Light group with M = 6.37, with this difference being statistically significant as well with a p-value of 0.003 (p < 0.01). The complete breakdown of NPS scores can be found in Appendix, Table 8. The detailed statistical results can be found in Appendix (Table 10).

### 6.2 Slow vs. Fast thinking

The results for the thinking mode questions indicate significant differences across the groups (Basic, Light, Full) regarding how participants perceived their thinking processes after reading an article. The results are visualized in Figure 7 and Figure 8. The variables that showed statistically significant differences among the groups were Speed, Processing, Control, and Adaptability. The detailed results can be found in Appendix (Table 9). Further, we first describe the overall result and then the result across individual thinking mode variables.

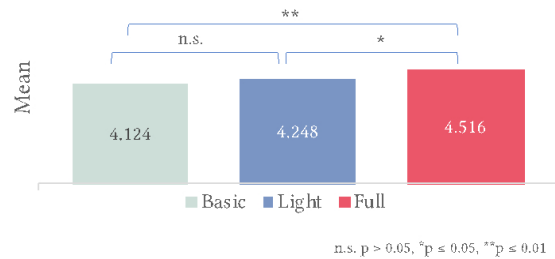

Figure 7: Thinking mode (overall)

**Thinking mode (overall):** The overall result for the thinking mode calculated as a mean of the scores for individual variables, indicates significant differences across the three groups. The Full group scored the highest in critical thinking (M = 4.516), followed by the Light group (M = 4.248) and the Basic group (M = 4.124). The ANOVA showed a statistically significant difference with a p-value of 0.02 (p < 0.05). Further, the t-tests revealed a significant difference between the Light and Full groups with a p-value of 0.013 (p < 0.05) and between the Basic and Full groups (p < 0.001), but not between the Basic and Light groups. This suggests that the treatment in the Full group was most effective in enhancing critical thinking.

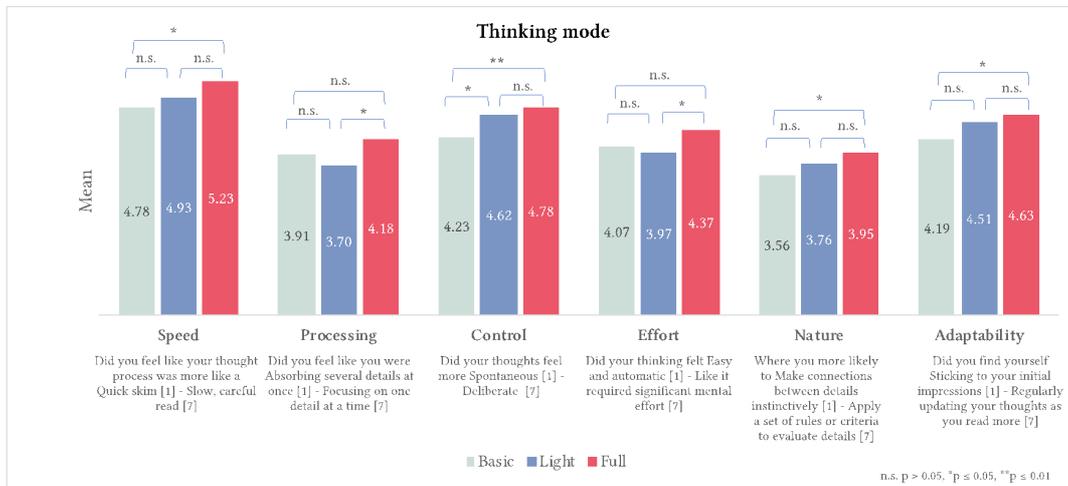

Figure 8: Thinking mode questions

**Speed**: When asked about their thinking speed, participants in the Full group reported the slowest reading speed with M = 5.23, followed by the Light group with M = 4.93 and the Basic group with M = 4.78. The ANOVA showed a p-value of 0.0285 (p < 0.05), indicating a statistically significant difference among the groups. The t-test between the Basic and Full groups revealed a significant difference with a p-value of 0.0115 (p < 0.01).

**Processing**: Participants in the Full group also reported focusing on one detail at a time as opposed to absorbing several details at once more than those in the Light and Basic groups, with M = 4.18, M = 3.70, and M = 3.91, respectively. The ANOVA p-value was 0.0498 (p < 0.05), suggesting a statistically significant difference among the groups. However, the t-tests between the Basic and Light groups and between the Basic and Full groups have no significant (n.s.) differences. The t-test between the Light and Full groups showed a p-value of 0.0172 (p < 0.05), indicating a significant difference. This may reflect that the Full version may encourage a more detailed content analysis than the Light version. This can be explained by the decrease in the average value from the Basic to the Light group.

**Control**: For the thought control, the Full group had a score of M = 4.78, followed by the Light group at M = 4.62 and the Basic group at M = 4.23. The ANOVA p-value was 0.0048 (p < 0.01), suggesting a highly significant difference among the groups. The t-test between the Basic and Light groups with a p-value of 0.0193 (p < 0.05) indicates a significant difference. The t-test between the Basic and Full groups has a highly significant p-value of 0.0017 (p < 0.01). This suggests that the Full and Light versions may promote more deliberate thinking.

**Effort**: For the mental effort, the Full group had the highest average score of M = 4.37, indicating they felt the participants' thinking required more significant mental effort. This was followed by the Basic group with an average of M = 4.07 and the Light group with M = 3.97. The ANOVA p-value was 0.0934 (n.s.). However, the t-test between the Light and Full groups showed a significant p-value of 0.041 (p < 0.05).

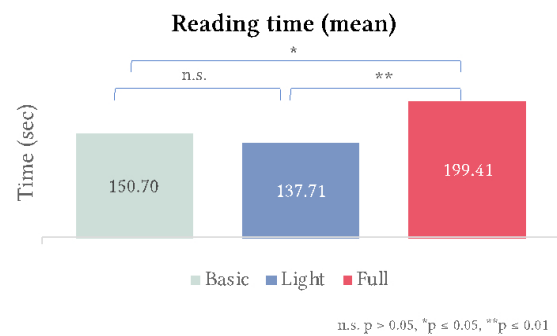

Figure 9: Reading time

**Nature**: Participants in the Full group were more likely to apply a set of rules or criteria to evaluate details, with an average score of M = 3.95, compared to the Light group's M = 3.76 and the Basic group's M = 3.56. The ANOVA p-value was 0.0834 (n.s.). However, the t-test between the Basic and Full groups has a p-value of 0.0275 (p < 0.05), which is significant.

**Adaptability**: Regarding the adaptability of the thinking process, the Full group scored the highest with M = 4.63, followed by the Light group at M = 4.51 and the Basic group at M = 4.19. The ANOVA p-value was 0.0423 (p < 0.05), indicating a statistically significant difference among the groups. The t-test between the Basic and Full groups had a p-value of 0.0169 (p < 0.05), which is significantly different.

## 6.3 Reading time

There is a difference between the mean reading time for the three different groups. Basic group had average reading time at M = 150.7 sec, Light at M = 137.71 sec, and Full at M = 199,41 sec. The results are visualized in Figure 9. The ANOVA yielded a statistically significant effect with a p-value of 0.0005 (p < 0.01), indicating that at least one of the groups differed significantly in mean reading time. Between Basic and Light groups, the t-test resulted in a p-value of 0.41, indicating a non-significant difference. Between Basic and Full groups, the t-test produced a p-value of 0.011 (p < 0.05), indicating a statistically significant difference in mean reading times. The t-test yielded a p-value of 0.000025 (p < 0.01) between the Light and Full groups, revealing a highly significant difference in mean reading times.

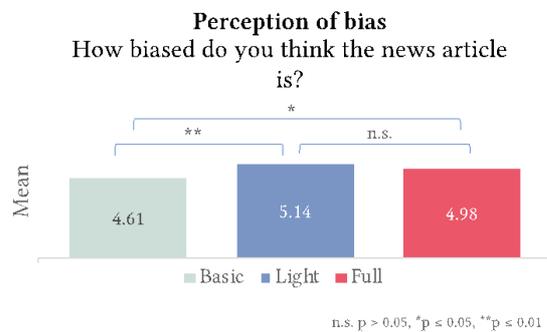

Figure 10: Perception of bias

### 6.4 Perception of bias in the news

Participants found the news articles more biased in the Light group (M = 5.14) and the Full group (M = 4.98) compared to the Basic group (M = 4.61). The ANOVA p-value is 0.0079 (p < 0.01), indicating that the difference in perceived bias among the groups is statistically significant. The t-test shows significant differences between the Basic and Light groups (p = 0.0025, p < 0.01) and between the Basic and Full groups (p = 0.0356, p < 0.05). The results are visualized in Figure 10.

### 6.5 Detection and explanation quality

The expert survey results show a high percentage of correctness in the propaganda detection of ClarifAI (Figure 11). 65% of the cases were rated as 'Good,' indicating that ClarifAI correctly identified propaganda according to both experts. 32% of cases where there was no agreement between the two experts. Only 3% of the cases were rated as 'Bad,' meaning ClarifAI's propaganda detection was incorrect, according to both experts. Despite the lack of agreement in a third of the cases, the positive ratings suggest that the tool is largely effective in correctly identifying propaganda.

The expert survey showed, in general, good or neutral results across different aspects of the explanations (Figure 12), confirming their general credibility. For accuracy, 15% of the responses were rated as bad, 35% as neutral, and 50% as good. Informativeness showed similar results, with 14% bad, 36% neutral, and 50% good. Believability and clarity scored higher, with 7% and 5% bad scores, respectively. These two aspects also received the most positive feedback, with 64% of the experts finding the explanations believable and 62% considering them clear and well-written. Table 11 shows more details on the expert evaluation for both the correctness of propaganda technique detection and the explanation quality.

### 6.6 Qualitative feedback

The *Light* version of ClarifAI sparked interest and '*added a different dimension to reading news articles,*' as one participant mentioned. However, users often expressed a need for more clarity and detailed explanations of why something was marked as propaganda. One participant formulated it here: '*There was no breakdown over WHY something was propaganda, so I'm not really sure I learned anything.*' Suggestions for improvement included offering alternative views from trustworthy sources and identifying the severity of propaganda techniques. Some concerns were raised about the trustworthiness of ClarifAI and the potential risks of politicization. For example, one participant mentioned: '*I distrust the notion of an arbiter which is used to suggest or determine propaganda, as this seems to inherently carry the risk of being politicized or weaponized.*' Despite these challenges, the light version was seen as a novel and engaging approach to propaganda detection.

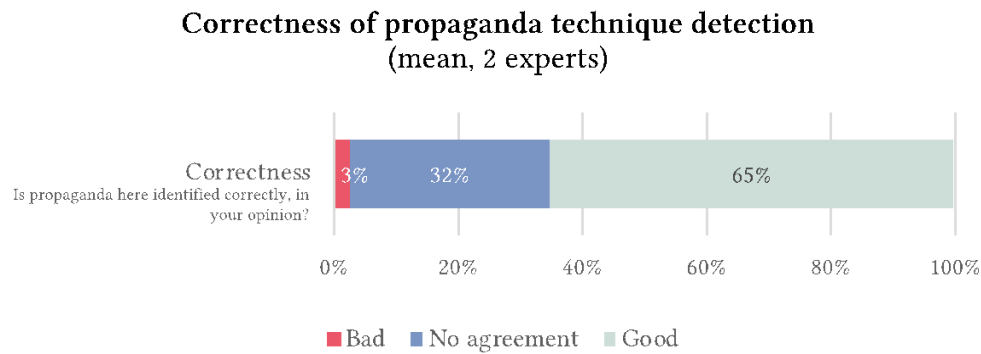

Figure 11: Expert assessment of the correctness of propaganda technique detection

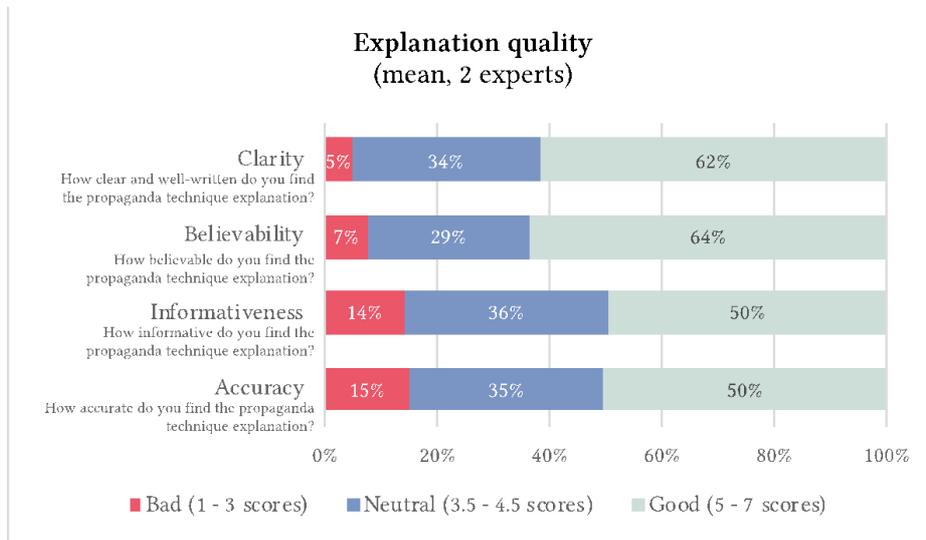

Figure 12: Expert assessment of the explanations' quality

The *Full* version of ClarifAI received even more positive comments from users, who found it helpful in understanding and analyzing propaganda content. Participants appreciated its perceived accuracy and the insight it provided into the context of articles. One participant suggested it can be useful in a discussion with friends prone to propaganda and fake news. Some suggested improvements included improving the visual aesthetics of highlights' colors to enhance the readability of texts. Pairing the tool with fact-checkers was suggested by the participants several times as a useful additional element. For example, one participant stated: '*I would find fact-checkers paired with this to be helpful, e.g., (...) When it says mentioning hospitals, schools, kindergartens is fear-mongering then say whether it was true or not or why they were located there, etc.*' Some concerns were also raised about perceived biases within the tool and possible inaccuracies in propaganda identification. For example, one participant mentioned the potential risk of inherent biases and that the quality of the detection of different techniques may vary: '*I think some bias from the creators may shine through on this tool. The last example was relevant info, for example - pointing out that a figure made a statement online then deleted it is not a mechanism of propaganda. I think the "doubt" and "appeal to fear / authority" were probably the most flawed measures but the rest were great.*' Overall, users wanted to continue using ClarifAI and recognized its value in helping them critically assess potential propaganda content.

### 6.7 Cost estimation

To run the LLM-generated propaganda detection, localization, and explanation, we estimate the cost of usage for the tokens of the GPT-4 API [85]. At the time of writing, the cost of tokens for the 8000 tokens context GPT-4 model is *$0.03 / 1000* tokens for the input and *$0.06 / 1000* tokens for the model's output. Further, the official documentation of OpenAI states that 100 tokens equal approximately 75 words [86]. It can be estimated that the input consisting of the detection prompt (758 tokens) and the input article (667 tokens) given an article length of 500 words, roughly leads to a cost of *(758 tokens +667 tokens) * $0.03 / 1000* tokens ≈ *$0.0427* cost for the input of the detection. Based on our experiments, we estimate that roughly 350 tokens are used for the output of the propaganda detection, i.e., *350 tokens * $0.06 / 1000* tokens = *$0.021*. This leads to a total cost of *$0.0637* for detecting propaganda. If, in the next step, each found technique will be located and explained, we utilize roughly 225 tokens for the localization and explanation prompt and again 667 tokens for the input article, leading to a cost for input token of *(225 tokens +667 tokens) * $0.03 / 1000* tokens ≈ *$0.0268*. Per an identified technique, the output consumed, on average, another 100 tokens, leading to a cost of *100 tokens * $0.06 / 1000* tokens = *$0.006*. To identify a single technique thus costs *$0.0328*. Therefore, if we approximate that three techniques can be found in an article, the total cost per article sums up to *$0.1621*.

## 7 DISCUSSION

Our study makes three key contributions we discuss in this section.

### 7.1 Context-specific digital nudging

One contribution of this study is the design of ClarifAI, a browser extension using GPT-4, which enacts digital nudges during the reading of textual content, itself. With this, we introduced a novel propaganda detection tool that employs context-specific digital nudges and the recent developments of LLMs to foster critical thinking in news consumers. Based on the literature, we carefully selected [20] and adopted five digital nudges to our case. These nudges altogether target three behavioral components outlined by Fogg's behavior model for persuasive design [57]: they simplify the task (of identifying propaganda in the news), increase motivation, and remind the task. Although the individual nudges, such as a default setting [20], [65] and the implemented features, such as a browser extension or text highlights [62], [63], [64], are not new in themselves, all together in combination they offer an effective tool while employing LLMs which provide reasoning to propaganda detection.

The proposed digital nudges (DN1 to DN5) function in a complementary manner and are synergetic. Each serves a specific purpose and contributes to the overall goal of enhancing critical thinking. For example, Immediate Insight (DN1) is an entry point and ensures users have an easy-to-use tool directly in their browsers. The design of ClarifAI also has the potential to address the issue of the typically short-lived positive impact of digital nudges [20]. However, the long-term effects should be further studied in a longitudinal study. By making ClarifAI consistently available in the user's browser (DN1 and DN2), we ensure that the nudges for critical thinking are continuously present and thus sustain their positive impact over time, thus mitigating the 'continued influence effect' [68], [69]. So, DN1 and DN2 create a basis for the other nudges to be effective. While DN2 and DN3 alert the user to potential propaganda, DN4 builds upon them and colors the content to act as a 'stumbling stone' that disrupts normal reading patterns. This makes users pause and reconsider the information. However, in line with prior observation, it is important to ensure that ClarifAI is not intrusive to the degree it gets ignored or turned off by the user. For example, this can be achieved using a visible but not too aggressive coloring scheme. Explainer (DN5) then serves as a closing loop, offering the user the 'what' and the 'why.' DN5 provides them with the knowledge they need to understand the manipulation techniques employed better. The nudges also have a temporal relationship: DN1 lays the groundwork, DN2 and DN3 maintain a constant state of alertness, DN4 provides occasional disruptions for more critical engagement, and DN5 offers additional insights that could have long-term benefits. We believe that the proposed combination of these nudges is especially effective and leaving one or another out may influence the intended outcome.

Reflecting on the effectiveness of our digital nudges, we are cautious, considering that research indicates only about 60% of nudges are effective [65]. In our study, the nudges were not evaluated individually. We argue that a combination of digital nudges might be necessary to achieve the desired outcome. In our application, we aimed to engage both System 1 (intuitive thinking) and System 2 (analytical thinking), but we did so through different nudges working in combination. Some nudges laid the groundwork for others, enhancing their effectiveness. For example, the Explainer nudge relies on prior alerts and detections. Without these, it would have no context for explanation. However, if there are multiple desired outcomes, this can become more complicated. In real-world scenarios, where multiple design objectives often coexist, a nudge effective for one specific goal might be counterproductive or even detrimental for another. Thus, we can conclude that a combination of different nudges can be beneficial. However, their choice needs to be carefully assessed based on the set design goals, intended outcomes, and potential conflicts between them.

Our research adds to the growing body of literature suggesting that the effectiveness of nudges can be context-dependent [20], [65], contributing to prior research about nudges in misinformation and fake news studies [55], [58], [59]. Our study also serves as a proof of concept, showing that automated propaganda detection tools can have real-world applications and benefit end users [9], [10], [11], [30]. Building on prior academic research, which focused on the technical aspects of ML models, we extend it with the user perspective, which is crucial for such tools to be successfully adopted.

### 7.2 Critical thinking

Second, our study is the first one to combine the dual system thinking perspective and the design of automated propaganda detection tools. This is particularly interesting because most of the prior work in propaganda detection [9], [10], [11], [30] does not address the psychological mechanisms that make propaganda effective or not. We suggest that understanding these mechanisms might lead to more effective use of such tools.

The evaluation results showed that the Full version of the tool was especially effective in slowing down reading speed and promoting detailed analysis, highlighting the design's usefulness for enhancing cognitive engagement. While the differences between the experimental groups are statistically significant, we see that the absolute differences between the means are not large. This brings us to the conclusion that *subtle* behavior change, as proposed by the concept of digital nudges, leads to a *subtle* change in critical thinking approach. Even subtle design elements, guided by dual system thinking, can be implemented to nudge users towards more informed and deliberate information consumption.

Feedback on the Full version of ClarifAI shows it can be valuable for both individual and collective critical thinking, encouraging users to challenge not only their perspectives but also those of people in their social circles. Participants specifically mentioned using the tool in social discussions. This suggests more versatility of the tool and its broader use, which should be studied further. For example, if users find the extension too intrusive, they might initially use it to vet a new media source and, upon developing trust, may not feel the need to verify every article or verify them only occasionally.

In our study, we primarily relied on dual system theory [19] as part of our design goals and the means of assessing critical thinking of news readers. This approach influenced both the tool's design and our evaluation setting, responding to the call for more interventions that influence critical thinking in news consumption [54]. In designing ClarifAI, we aimed to balance the strengths of both intuitive (System 1) and deliberative (System 2) thinking. This meant facilitating both quick, intuitive interactions and supporting more reflective thinking through provided explanations simultaneously, as reflected by our design goals. This also addressed some of the existing critiques [51] of the dual system theory as labeling certain ways of thinking and the resulting biases as good or bad [52]. However, it is important to acknowledge that the influence of dual system theory on our design primarily occurs at a conceptual level, recognizing the need for both intuitive and deliberate thinking. However, in practical terms, the literature on nudging provided more tangible benefits for our design process. This reflects a situation where a theory provides a broad framework, while specific design choices are more directly shaped by applied approaches like nudging.

Regarding the application of dual system theory in assessing critical thinking during our evaluation, we found it beneficial to use this framework for multiple reasons. To understand their thinking mode, we asked participants to choose on the spectrum between two distinct options: one from System 1 and one from System 2. Opting for a single-item question about critical thinking could lead to biased responses, as most people will likely affirm that they think critically. It is uncommon for individuals to admit a lack of critical thinking. Furthermore, merely posing a question about critical thinking can potentially prompt participants to engage in more critical reflection than they might otherwise. We intentionally did not analyze relationships between individual variables to avoid complicating the analysis and its interpretation, as our research aim was not to explore these relationships. Nonetheless, our design influenced various aspects of thinking processes, such as speed and control. For future research, we believe that other evaluation methods, such as examining user behavior via eye-tracking or mouse tracking and neuroscience techniques to measure brain activity, could enrich and complement our findings.

However, recognizing the critiques of Kahneman's dual-system theory [51], like the oversimplification of thinking cognition processes, we understand that it might affect how ClarifAI works and how we interpret our results. Our tool aims to foster critical thinking by influencing both fast, intuitive reactions and slower, more deliberate thought. Yet, the real process of thinking is more complex and intertwined than this division suggests. So, ClarifAI is not the final answer but a step towards better news understanding. We suggest that for future work, such design, rooted in dual-system theory, needs to adapt as we learn more about how people process information and as the world of digital news evolves.

Moreover, our research also motivates further discussion on who may be a potential user of such propaganda detection tools. Users who are already critical thinkers may not find such a tool necessary, while those deeply entrenched in propaganda are unlikely to trust a tool that challenges their established beliefs [50]. However, such a tool may be most relevant for individuals between these extremes. Like in political campaigns, the goal may be to convert this potentially large non-expert group into more critical thinkers. However, it is also necessary to recognize the ethical implications of the influence of such tools when it comes to 'the right to be wrong.' Even when we aim to foster more informed decisions, the decision to accept or reject information should still stay with the individual user.

### 7.3 Value of explanations

Third, the results of our study demonstrate the value of explanations accompanying propaganda detection. While the Light version of propaganda detection, which merely highlights propaganda instances, influences how readers perceive bias, it does little to encourage a more critical approach to thinking. In some cases, like fostering a sense of control over one's thinking, the Light version even led to quicker, less critical thought processes. This might be attributed to the rapid reactions that people tend to have in response to visually prominent colors. However, the increase in perceived bias of the overall article could be explained by the reader's impression that the presence of any highlights in the text implies the article is inherently biased. So, we conclude that the change in critical thinking can be attributed to explanations provided by the Full version. Unlike the highlights in the Light version, the explanations provide additional content for the reader. This may trigger the shift towards more critical thinking. Furthermore, the study's participants explicitly asked for explanations that would accompany such detections, indicating the potential usefulness and educational effect of these explanations.

Additionally, our expert evaluation showed that the correctness of the propaganda technique identification and the quality of generated explanations are quite high, though not perfect. This can be explained by several factors, including the subjectiveness of the task and the LLMs' propensity to hallucinate [38]. Even if the LLM-generated explanations are imperfect, they can still prompt users to pause and think critically. Given the complexity of identifying propaganda, this pause for reflection can be invaluable in fostering more analytical thinking. However, in high-stakes scenarios like medical AI, where both complexity and the need for precision are present, such limitations may carry serious implications. Finally, when comparing the evaluation scores, LLM-based explanations score higher in clarity and believability than in accuracy and informativeness. This indicates that while LLMs are effective at creating user-friendly explanations, there is a need for additional research to ensure these explanations do not inadvertently perpetuate misinformation of another kind or political biases inherent in LLMs [33], [34]. There is a potential for algorithmic bias to influence what is identified as propaganda and how it is explained, which could inadvertently shape users' perceptions of news credibility.

Tools such as proposed ClarifAI also raise socio-technical considerations. While it aims to foster critical engagement with digital content, there is a risk that users may become overly reliant on automated suggestions of the propaganda detection tool. This may potentially undermine independent critical thinking skills. While prior research indicated that explanations can reduce overreliance on AI [39], overreliance on explanations may become an issue and may potentially undermine independent critical thinking skills, especially when AI tools work in an 'always on' mode. Moreover, the broader implications of algorithmic bias and the ethical use of automated propaganda detection tools call for further research to understand whether or not they contribute positively to public discourse and democracy.

## 8 CONCLUSIONS AND LIMITATIONS

Based on the results of our study, we conclude that we could research the goals initially set, such as promoting a shift from reflexive to analytical thinking, thereby counteracting the negative impacts of propagandistic content on humans' critical thinking. Our study has several limitations worth noting. Firstly, the online experiment relies on self-reported results from participants, introducing a layer of subjectivity. Future work could employ less subjective assessment methods like eye-tracking studies. This will provide further empirical evidence for the effectiveness of digital nudging and its impact on the critical thinking approach. Secondly, despite their expertise, we know expert evaluations may also carry inherent biases. Additionally, our research does not examine the long-term efficacy or potential issues arising from prolonged use of ClarifAI;

future studies could focus on these aspects. Furthermore, our study concentrates solely on the context of digital news. However, we believe that our solution may have an impact beyond, for example, in a forum or social media contexts where the dissemination of propaganda is especially prevalent.

Thirdly, the cost of such a system like ClarifAI needs to be considered as it may be a practical limitation of ClarifAI. For our implementation, we did a rough cost estimation of *$0.1621* per article for using the GPT-4 API, with three techniques being found per article. Still, if an open-source model is deployed, the cost would vary based on the employed GPU, its electricity consumption, and the price of electricity in the location where the model is hosted. This approach may reduce per-analysis costs and align with the varying budget constraints of potential users. In our project, the question of funding the operational costs of ClarifAI remains open. While it is unlikely that end users would bear the costs for intensive daily use because of potentially high expenses, we are exploring several possibilities. One possibility is a freemium model, offering basic functionality or a set number of API calls for free, with advanced features available for a fee. Additionally, we are considering a business-to-business financing model targeting organizations interested in analyzing and reporting on propaganda and misinformation. These organizations could benefit from automated propaganda reports generated by ClarifAI. However, we acknowledge that the feasibility of these models needs further evaluation by our research team in future work.

Fourthly, another limitation of the study relates to using a controlled experiment with Figma prototypes for testing ClarifAI rather than allowing participants to use it in their natural online environments with familiar news sources, potentially affecting the study's ecological validity. To address this, future research should involve deploying ClarifAI directly into users' everyday online news consumption settings, allowing for interaction with the news sources they typically consume. This approach will provide insights into how readers' biases and long-term usage affect the tool's effectiveness and consequences, both intended and not intended, such as overreliance on ClarifAI's suggestions or turning it completely off. Such naturalistic testing could reveal different challenges and patterns of use, which are important for refining ClarifAI's design in real-world scenarios. Lastly, to contribute to more stable democracies, further interdisciplinary research, including social, communication, and political sciences, should explore the impacts of such tools on democratic institutions and processes.

## ACKNOWLEDGMENTS

To improve the readability and coherence, and to correct grammatical errors in this text, we have employed AI tools such as ChatGPT and Grammarly. Otherwise, all ideas and findings presented in the work belong to the authors. We have thoroughly reviewed and refined the AI-generated content to ensure its accuracy. We assume full responsibility for the final content presented in this work.

We would like to thank our participants and the anonymous associate editors and reviewers for their valuable comments and helpful suggestions. We also thank the Digital Society Initiative of the University of Zurich and the Digitalization Initiative of the Zurich Higher Education Institutions (DIZH) for partially financing this study under the DIZH postdoc fellowship of Liudmila Zavolokina.

## REFERENCES

[1] R. Paul and L. Elder, How to Detect Media Bias & Propaganda: In the National and World News: Based on the Critical Conepts & Tools. Foundation for Critical Thinking, 2008.

[2] G. Da San Martino, S. Shaar, Y. Zhang, S. Yu, A. Barrón-Cedeno, and P. Nakov, "Prta: A system to support the analysis of propaganda techniques in the news," in Proceedings of the 58th Annual Meeting of the Association for Computational Linguistics: System Demonstrations, 2020, pp. 287–293.

[3] G. Da San Martino, S. Cresci, A. Barrón-Cedeño, S. Yu, R. Di Pietro, and P. Nakov, "A survey on computational propaganda detection," arXiv preprint arXiv:2007.08024, 2020.

[4] C. Paul and M. Matthews, "The Russian 'firehose of falsehood' propaganda model," Rand Corporation, vol. 2, no. 7, pp. 1–10, 2016.

[5] M. Alyukov, "Propaganda, authoritarianism and Russia's invasion of Ukraine," Nature Human Behaviour, vol. 6, no. 6, pp. 763–765, 2022.

[6] "Overview and key findings of the 2022 Digital News Report," Reuters Institute for the Study of Journalism. Accessed: Aug. 07, 2023. [Online]. Available: https://reutersinstitute.politics.ox.ac.uk/digital-news-report/2022/dnr-executive-summary

[7] D. Kahneman, "Two systems in the mind," Bulletin of the American Academy of Arts and Sciences, vol. 65, no. 2, pp. 55–59, 2012.

[8] D. Kahneman, "Maps of bounded rationality: A perspective on intuitive judgement and choice," 2002.

[9] M. L. Della Vedova, E. Tacchini, S. Moret, G. Ballarin, M. DiPierro, and L. De Alfaro, "Automatic online fake news detection combining content and social signals," in 2018 22nd conference of open innovations association (FRUCT), IEEE, 2018, pp. 272–279.

[10] E. Qawasmeh, M. Tawalbeh, and M. Abdullah, "Automatic identification of fake news using deep learning," in 2019 Sixth international conference on social networks analysis, Management and Security (SNAMS), IEEE, 2019, pp. 383–388.

[11] A. A. A. Ahmed, A. Aljabouh, P. K. Donepudi, and M. S. Choi, "Detecting fake news using machine learning: A systematic literature review," arXiv preprint arXiv:2102.04458, 2021.

[12] A. Radford, K. Narasimhan, T. Salimans, I. Sutskever, and others, "Improving language understanding by generative pre-training," 2018.

[13] A. Radford et al., "Language models are unsupervised multitask learners," OpenAI blog, vol. 1, no. 8, p. 9, 2019.

[14] T. Brown et al., "Language models are few-shot learners," Advances in neural information processing systems, vol. 33, pp. 1877–1901, 2020.

[15] OpenAI, "GPT-4 Technical Report." 2023. Accessed: Mar. 16, 2023. [Online]. Available: https://cdn.openai.com/papers/gpt-4.pdf

[16] J. Devlin, M.-W. Chang, K. Lee, and K. Toutanova, "Bert: Pre-training of deep bidirectional transformers for language understanding," arXiv preprint arXiv:1810.04805, 2018.

[17] H. Touvron et al., "Llama: Open and efficient foundation language models," arXiv preprint arXiv:2302.13971, 2023.

[18] K. Sprenkamp, D. G. Jones, and L. Zavolokina, "Large Language Models for Propaganda Detection." arXiv, Nov. 27, 2023. doi: 10.48550/arXiv.2310.06422.

[19] D. Kahneman, Thinking, fast and slow. macmillan, 2011.

[20] A. Caraban, E. Karapanos, D. Gonçalves, and P. Campos, "23 ways to nudge: A review of technology-mediated nudging in human-computer interaction," in Proceedings of the 2019 CHI conference on human factors in computing systems, 2019, pp. 1–15.

[21] K. Peffers, T. Tuunanen, M. A. Rothenberger, and S. Chatterjee, "A design science research methodology for information systems research," Journal of management information systems, vol. 24, no. 3, pp. 45–77, 2007.

[22] J. F. Nunamaker Jr, R. O. Briggs, D. C. Derrick, and G. Schwabe, "The last research mile: Achieving both rigor and relevance in information systems research," Journal of Management Information Systems, vol. 32, no. 3, pp. 10–47, 2015.

[23] "Главная," StopFake. Accessed: Mar. 26, 2023. [Online]. Available: https://www.stopfake.org

[24] "bellingcat - the home of online investigations," bellingcat. Accessed: Mar. 26, 2023. [Online]. Available: https://www.bellingcat.com/

[25] "Our Mission," FactCheck.org. Accessed: Mar. 26, 2023. [Online]. Available: https://www.factcheck.org/about/our-mission/

[26] "PolitiFact." Accessed: Mar. 26, 2023. [Online]. Available: https://www.politifact.com/

[27] J. Piskorski, N. Stefanovitch, G. Da San Martino, and P. Nakov, "Semeval-2023 task 3: Detecting the category, the framing, and the persuasion techniques in online news in a multi-lingual setup," in Proceedings of the 17th International Workshop on Semantic Evaluation (SemEval-2023), 2023, pp. 2343–2361.

[28] H. Rashkin, E. Choi, J. Y. Jang, S. Volkova, and Y. Choi, "Truth of varying shades: Analyzing language in fake news and political fact-checking," in Proceedings of the 2017 conference on empirical methods in natural language processing, 2017, pp. 2931–2937.

[29] A. Barrón-Cedeno, I. Jaradat, G. Da San Martino, and P. Nakov, "Proppy: Organizing the news based on their propagandistic content," Information Processing & Management, vol. 56, no. 5, pp. 1849–1864, 2019.
[30] G. Da San Martino, S. Yu, A. Barrón-Cedeno, R. Petrov, and P. Nakov, "Fine-grained analysis of propaganda in news article," in Proceedings of the 2019 conference on empirical methods in natural language processing and the 9th international joint conference on natural language processing (EMNLP-IJCNLP), 2019, pp. 5636–5646.
[31] G. Da San Martino, A. Barrón-Cedeno, H. Wachsmuth, R. Petrov, and P. Nakov, "SemEval-2020 task 11: Detection of propaganda techniques in news articles," arXiv preprint arXiv:2009.02696, 2020.
[32] S. Yu, G. D. S. Martino, M. Mohtarami, J. Glass, and P. Nakov, "Interpretable propaganda detection in news articles," arXiv preprint arXiv:2108.12802, 2021.
[33] E. M. Bender, T. Gebru, A. McMillan-Major, and S. Shmitchell, "On the dangers of stochastic parrots: Can language models be too big?," in Proceedings of the 2021 ACM conference on fairness, accountability, and transparency, 2021, pp. 610–623.
[34] S. Feng, C. Y. Park, Y. Liu, and Y. Tsvetkov, "From Pretraining Data to Language Models to Downstream Tasks: Tracking the Trails of Political Biases Leading to Unfair NLP Models," arXiv preprint arXiv:2305.08283, 2023.
[35] D. Dimitrov et al., "Detecting propaganda techniques in memes," arXiv preprint arXiv:2109.08013, 2021.
[36] P. Nakov, F. Alam, S. Shaar, G. Da San Martino, and Y. Zhang, "COVID-19 in Bulgarian social media: Factuality, harmfulness, propaganda, and framing," in Proceedings of the International Conference on Recent Advances in Natural Language Processing (RANLP 2021), 2021, pp. 997–1009.
[37] A. Abedalla, A. Al-Sadi, and M. Abdullah, "A closer look at fake news detection: A deep learning perspective," in Proceedings of the 3rd International Conference on Advances in Artificial Intelligence, 2019, pp. 24–28.
[38] Y. Bang et al., "A multitask, multilingual, multimodal evaluation of chatgpt on reasoning, hallucination, and interactivity," arXiv preprint arXiv:2302.04023, 2023.
[39] H. Vasconcelos, M. Jörke, M. Grunde-McLaughlin, T. Gerstenberg, M. S. Bernstein, and R. Krishna, "Explanations can reduce overreliance on ai systems during decision-making," Proceedings of the ACM on Human-Computer Interaction, vol. 7, no. CSCW1, pp. 1–38, 2023.
[40] B. Tafur and A. Sarkar, "User Perceptions of Automatic Fake News Detection: Can Algorithms Fight Online Misinformation?," arXiv preprint arXiv:2304.07926, 2023.
[41] G. Pennycook, A. Bear, E. T. Collins, and D. G. Rand, "The implied truth effect: Attaching warnings to a subset of fake news headlines increases perceived accuracy of headlines without warnings," Management science, vol. 66, no. 11, pp. 4944–4957, 2020.
[42] F. Jahanbakhsh, Y. Katsis, D. Wang, L. Popa, and M. Muller, "Exploring the Use of Personalized AI for Identifying Misinformation on Social Media," in Proceedings of the 2023 CHI Conference on Human Factors in Computing Systems, 2023, pp. 1–27.
[43] Z. Epstein, A. J. Berinsky, R. Cole, A. Gully, G. Pennycook, and D. G. Rand, "Developing an accuracy-prompt toolkit to reduce COVID-19 misinformation online," 2021, Accessed: Dec. 12, 2023. [Online]. Available: https://dspace.mit.edu/handle/1721.1/138124
[44] G. Pennycook, J. McPhetres, Y. Zhang, J. G. Lu, and D. G. Rand, "Fighting COVID-19 Misinformation on Social Media: Experimental Evidence for a Scalable Accuracy-Nudge Intervention," Psychol Sci, vol. 31, no. 7, pp. 770–780, Jul. 2020, doi: 10.1177/0956797620939054.
[45] F. Jahanbakhsh, A. X. Zhang, A. J. Berinsky, G. Pennycook, D. G. Rand, and D. R. Karger, "Exploring Lightweight Interventions at Posting Time to Reduce the Sharing of Misinformation on Social Media," Proc. ACM Hum.-Comput. Interact., vol. 5, no. CSCW1, pp. 1–42, Apr. 2021, doi: 10.1145/3449092.
[46] G. Pennycook, Z. Epstein, M. Mosleh, A. A. Arechar, D. Eckles, and D. G. Rand, "Shifting attention to accuracy can reduce misinformation online," Nature, vol. 592, no. 7855, pp. 590–595, 2021.
[47] A. Das, H. Liu, V. Kovatchev, and M. Lease, "The state of human-centered NLP technology for fact-checking," Information processing & management, vol. 60, no. 2, p. 103219, 2023.
[48] V. Achimescu and D. Sultanescu, "Feeding the troll detection algorithm: Informal flags used as labels in classification models to identify perceived computational propaganda," First Monday, 2020.
[49] F. Strack and R. Deutsch, "Reflective and impulsive determinants of social behavior," Personality and social psychology review, vol. 8, no. 3, pp. 220–247, 2004.
[50] D. Kahneman, J. L. Knetsch, and R. H. Thaler, "Anomalies: The endowment effect, loss aversion, and status quo bias," Journal of Economic perspectives, vol. 5, no. 1, pp. 193–206, 1991.
[51] J. St. B. T. Evans and K. E. Stanovich, "Dual-Process Theories of Higher Cognition: Advancing the Debate," Perspect Psychol Sci, vol. 8, no. 3, pp. 223–241, May 2013, doi: 10.1177/1745691612460685.
[52] P. B. Vranas, "Gigerenzer's normative critique of Kahneman and Tversky," Cognition, vol. 76, no. 3, pp. 179–193, 2000.
[53] B. D. Horne, M. Gruppi, and S. Adali, "Trustworthy misinformation mitigation with soft information nudging," in 2019 first IEEE international conference on trust, privacy and security in intelligent systems and applications (TPS-ISA), IEEE, 2019, pp. 245–254.
[54] D. Pehlivanoglu, T. Lin, F. Deceus, A. Heemskerk, N. C. Ebner, and B. S. Cahill, "The role of analytical reasoning and source credibility on the evaluation of real and fake full-length news articles," Cogn. Research, vol. 6, no. 1, Art. no. 1, Dec. 2021, doi: 10.1186/s41235-021-00292-3.
[55] C. Thornhill, Q. Meeus, J. Peperkamp, and B. Berendt, "A digital nudge to counter confirmation bias," Frontiers in big data, vol. 2, p. 11, 2019.
[56] P. G. Hansen and A. M. Jespersen, "Nudge and the manipulation of choice: A framework for the responsible use of the nudge approach to behaviour change in public policy," European Journal of Risk Regulation, vol. 4, no. 1, pp. 3–28, 2013.
[57] B. J. Fogg, "A behavior model for persuasive design," in Proceedings of the 4th international conference on Persuasive Technology, 2009, pp. 1–7.
[58] L. Konstantinou, A. Caraban, and E. Karapanos, "Combating misinformation through nudging," in Human-Computer Interaction–INTERACT 2019: 17th IFIP TC 13 International Conference, Paphos, Cyprus, September 2–6, 2019, Proceedings, Part IV 17, Springer, 2019, pp. 630–634.
[59] A. Caraban, L. Konstantinou, and E. Karapanos, "The nudge deck: A design support tool for technology-mediated nudging," in Proceedings of the 2020 ACM Designing Interactive Systems Conference, 2020, pp. 395–406.
[60] K. J. K. Feng, N. Ritchie, P. Blumenthal, A. Parsons, and A. X. Zhang, "Examining the Impact of Provenance-Enabled Media on Trust and Accuracy Perceptions," Proc. ACM Hum.-Comput. Interact., vol. 7, no. CSCW2, pp. 1–42, Sep. 2023, doi: 10.1145/3610061.
[61] F. Jahanbakhsh, A. X. Zhang, and D. R. Karger, "Leveraging Structured Trusted-Peer Assessments to Combat Misinformation," Proc. ACM Hum.-Comput. Interact., vol. 6, no. CSCW2, pp. 1–40, Nov. 2022, doi: 10.1145/3555637.
[62] R. Ennals, B. Trushkowsky, and J. M. Agosta, "Highlighting disputed claims on the web," in Proceedings of the 19th international conference on World wide web, Raleigh North Carolina USA: ACM, Apr. 2010, pp. 341–350. doi: 10.1145/1772690.1772726.
[63] F. Jahanbakhsh, A. X. Zhang, K. Karahalios, and D. R. Karger, "Our Browser Extension Lets Readers Change the Headlines on News Articles, and You Won't Believe What They Did!," Proc. ACM Hum.-Comput. Interact., vol. 6, no. CSCW2, pp. 1–33, Nov. 2022, doi: 10.1145/3555643.
[64] M. M. U. Rony, N. Hassan, and M. Yousuf, "BaitBuster: a clickbait identification framework," in Proceedings of the AAAI Conference on Artificial Intelligence, 2018. Accessed: Dec. 12, 2023. [Online]. Available: https://ojs.aaai.org/index.php/AAAI/article/view/11378
[65] D. Hummel and A. Maedche, "How effective is nudging? A quantitative review on the effect sizes and limits of empirical nudging studies," Journal of Behavioral and Experimental Economics, vol. 80, pp. 47–58, 2019.
[66] G. S. Jowett and V. O'donnell, Propaganda & persuasion. Sage publications, 2018.
[67] R. S. Nickerson, "Confirmation bias: A ubiquitous phenomenon in many guises," Review of general psychology, vol. 2, no. 2, pp. 175–220, 1998.
[68] S. Lewandowsky, U. K. Ecker, C. M. Seifert, N. Schwarz, and J. Cook, "Misinformation and its correction: Continued influence and successful debiasing," Psychological science in the public interest, vol. 13, no. 3, pp. 106–131, 2012.
[69] U. K. H. Ecker et al., "The psychological drivers of misinformation belief and its resistance to correction," Nat Rev Psychol, vol. 1, no. 1, Art. no. 1, Jan. 2022, doi: 10.1038/s44159-021-00006-y.
[70] L. Hasher, D. Goldstein, and T. Toppino, "Frequency and the conference of referential validity," Journal of verbal learning and verbal behavior, vol. 16, no. 1, pp. 107–112, 1977.
[71] T. Brader, "Striking a responsive chord: How political ads motivate and persuade voters by appealing to emotions," American Journal of Political Science, vol. 49, no. 2, pp. 388–405, 2005.
[72] E. Bakshy, S. Messing, and L. A. Adamic, "Exposure to ideologically diverse news and opinion on Facebook," Science, vol. 348, no. 6239, pp. 1130–1132, 2015.
[73] "Prolific · Quickly find research participants you can trust." Accessed: Jul. 20, 2023. [Online]. Available: https://www.prolific.co/
[74] E. Peer, D. Rothschild, A. Gordon, Z. Evernden, and E. Damer, "Data quality of platforms and panels for online behavioral research," Behavior Research Methods, p. 1, 2022.
[75] G. Lins de Holanda Coelho, P. HP Hanel, and L. J. Wolf, "The very efficient assessment of need for cognition: Developing a six-item version," Assessment, vol. 27, no. 8, pp. 1870–1885, 2020.
[76] "Report: RT and Sputnik's Role in Russia's Disinformation and Propaganda Ecosystem," United States Department of State. Accessed: Aug. 27, 2023. [Online]. Available: https://www.state.gov/report-rt-and-sputniks-role-in-russias-disinformation-and-propaganda-ecosystem/
[77] "InfoWars," Wikipedia. Aug. 18, 2023. Accessed: Aug. 27, 2023. [Online]. Available: https://en.wikipedia.org/w/index.php?title$=$InfoWars&oldid$=$1170917917
[78] "RT - Breaking News, Russia News, World News and Video," RT International. Accessed: Aug. 27, 2023. [Online]. Available: https://www.rt.com

[79] "Infowars: There's a War on For Your Mind!" Accessed: Aug. 27, 2023. [Online]. Available: https://www.infowars.com/
[80] "Qualtrics integration guide," Prolific. Accessed: Aug. 30, 2023. [Online]. Available: https://researcher-help.prolific.co/hc/en-gb/articles/360009224113-Qualtrics-integration-guide
[81] Qualtrics, "Using Attention Checks in Your Surveys May Harm Data Quality," Qualtrics. Accessed: Aug. 30, 2023. [Online]. Available: https://www.qualtrics.com/blog/attention-checks-and-data-quality/
[82] M. Solnyshkina, R. Zamaletdinov, L. Gorodetskaya, and A. Gabitov, "Evaluating text complexity and Flesch-Kincaid grade level," Journal of social studies education research, vol. 8, no. 3, pp. 238–248, 2017.
[83] "Qualtrics XM - Experience Management Software," Qualtrics. Accessed: Jul. 21, 2023. [Online]. Available: https://www.qualtrics.com/uk/
[84] F. F. Reichheld, "The one number you need to grow," Harvard business review, vol. 81, no. 12, pp. 46–55, 2003.
[85] "Pricing." Accessed: Dec. 10, 2023. [Online]. Available: https://openai.com/pricing
[86] "What are tokens and how to count them? | OpenAI Help Center." Accessed: Dec. 12, 2023. [Online]. Available: https://help.openai.com/en/articles/4936856-what-are-tokens-and-how-to-count-them
[87] M. Abdullah, O. Altiti, and R. Obiedat, "Detecting propaganda techniques in english news articles using pre-trained transformers," in 2022 13th International Conference on Information and Communication Systems (ICICS), IEEE, 2022, pp. 301–308.

# A APPENDIX

Table 4: Propaganda techniques (based on [30], [87])

| Propaganda Technique | Definition | Example |
| --- | --- | --- |
| Appeal to Authority | Supposes that a claim is true because a valid authority or expert on the issue supports it | "The World Health Organisation stated, the new medicine is the most effective treatment for the disease." |
| Appeal to fear-prejudice | Builds support for an idea by instilling anxiety and/or panic in the audience towards an alternative | "Stop those refugees; they are terrorists." |
| Bandwagon, Reductio ad hitlerum | Justify actions or ideas because everyone else is doing it, or reject them because it's favored by groups despised by the target audience | "Would you vote for Clinton as president? 57% say yes." |
| Black-and-White Fallacy | Gives two alternative options as the only possibilities, when actually more options exist | "You must be a Republican or Democrat" |
| Causal Oversimplification | Assumes a single reason for an issue when there are multiple causes | "If France had not declared war on Germany, World War II would have never happened." |
| Doubt | Questioning the credibility of someone or something | "Is he ready to be the Mayor?" |
| Exaggeration, Minimisation | Either representing something in an excessive manner or making something seem less important than it actually is | "I was not fighting with her; we were just playing." |
| Flag-Waving | Playing on strong national feeling (or with respect to a group, e.g., race, gender, political preference) to justify or promote an action or idea | "Entering this war will make us have a better future in our country." |
| Loaded Language | Uses specific phrases and words that carry strong emotional impact to affect the audience | "A lone lawmaker's childish shouting." |
| Name Calling, Labeling | Gives a label to the object of the propaganda campaign as either the audience hates or loves | "Bush the Lesser." |
| Repetition | Repeats the message over and over in the article so that the audience will accept it | "Our great leader is the epitome of wisdom. Their decisions are always wise and just." |
| Slogans | A brief and striking phrase that contains labeling and stereotyping | "Make America great again!" |
| Thought-terminating Cliches | Words or phrases that discourage critical thought and useful discussion about a given topic | "It is what it is" |
| Whataboutism, Straw Men, Red Herring | Attempts to discredit an opponent's position by charging them with hypocrisy without directly disproving their argument | "They want to preserve the FBI's reputation." |

Table 5: Prompts employed

| Task | Prompt |
| --- | --- |
| **IF3. Real-time LLM-based detection** | You are a Text Classifier identifying 14 Propaganda Techniques within Newspaper Articles. These are the 14 propaganda techniques you classify with definitions and examples:<br><br>Appeal_to_fear-prejudice - Builds support for an idea by instilling anxiety and/or panic in the audience towards an alternative, e.g., 'Stop those refugees; they are terrorists.'<br>Appeal_to_Authority - Supposes that a claim is true because a valid authority or expert on the issue supports it, e.g., 'The World Health Organization stated the new medicine is the most effective treatment for the disease.'<br>Bandwagon, Reductio_ad_hitlerum - Justify actions or ideas because everyone else is doing it, or reject them because it's favored by groups despised by the target audience, e.g., "Would you vote for Clinton as president? 57% say yes."<br><br>[...]<br>*[for the full list of propaganda techniques, definitions, and examples, refer to Table 4]*<br><br>For the given article, please state which of the 14 propaganda techniques are present and give an explanation of why the technique is present in the article. If no propaganda technique was identified, return "no propaganda detected". An example output would list the propaganda techniques, with each technique on a new line, e.g.:<br>Loaded_Language - Your explanation of why this technique is present in the article.<br>Thought-terminating_Cliches - Your explanation of why this technique is present in the article.<br>Repetition - Your explanation of why this technique is present in the article.<br><br>Here is the article:<br>**{input_article}** |
| **IF5. LLM-generated explanations** | Please check the following article for the propaganda technique **{technique}**, defined as **{definition_of_technique}**. Then return the passage in the article that contains the propaganda technique, this can be a sentence, multiple setences, sub-sentence or words. Further, reason why you think the propaganda technique is present in the marked text passage.<br><br>The output should ONLY be in the following format:<br><propaganda technique>\|<passage in the article>\|<reason for the propaganda technique><br>For example:<br>Exaggeration\|Europe's own arsenals are so depleted, the Salvation Army could probably march upon Paris and conquer mainland Europe without a shot being fired.\|The author exaggerates the state of Europe's arsenals to make it seem as if they are extremely weak and depleted, suggesting that even a non-military organization like the Salvation Army could easily conquer Europe. This exaggeration is used to emphasize the desperation of Ukraine and its NATO backers.<br><br>Here is the article:<br>**{input_article}** |

Table 6: Demographic data of the experiment's participant

|  | Basic | Light | Full | Total |
|---|---|---|---|---|
| **Participants** | | | | |
| N | 81 (33%) | 84 (34%) | 83 (33%) | 248 |
| **Age** | | | | |
| 18 to 24 | 39 | 44 | 27 | 110 |
| 25 to 44 | 34 | 38 | 48 | 120 |
| 45 to 64 | 8 | 2 | 7 | 17 |
| 65 and over | - | - | 1 | 1 |
| **Gender** | | | | |
| Female | 40 | 41 | 42 | 123 |
| Male | 41 | 41 | 41 | 123 |
| Non-binary | - | 2 | - | 2 |
| **Education** | | | | |
| Less than high school diploma | - | - | 1 | 1 |
| High school diploma or equivalent | 9 | 6 | 9 | 24 |
| Some college/university, no degree | 24 | 34 | 11 | 69 |
| Associate degree | 2 | 6 | 5 | 13 |
| Bachelor's degree | 30 | 29 | 35 | 94 |
| Graduate degree (Master's or PhD) | 16 | 9 | 22 | 47 |
| **Country of residence** | | | | |
| United Kingdom | 42 | 43 | 45 | 130 |
| South Africa | 24 | 28 | 28 | 80 |
| Ireland | 5 | 1 | 1 | 7 |
| Canada | - | 6 | - | 6 |
| United States | 1 | - | 2 | 3 |
| Australia | 1 | - | 2 | 3 |
| Other countries | 7 | 5 | 3 | 15 |
| **Professional background** | | | | |
| Accounting, Finance | 3 | 9 | 7 | 19 |
| Arts | 2 | 9 | 3 | 14 |
| Biology | 4 | 6 | 6 | 16 |
| Business | 4 | - | 5 | 9 |
| Computer Science | 13 | 12 | 6 | 31 |
| Economics | 3 | 4 | 3 | 10 |
| Education | 4 | 3 | 4 | 11 |
| Engineering | 9 | 3 | 7 | 19 |
| Languages | 4 | 9 | 3 | 16 |
| Law | 3 | 3 | 5 | 11 |
| Marketing | 2 | 2 | 4 | 8 |
| Mathematics | 3 | 4 | 2 | 9 |
| Medicine | 5 | 5 | 5 | 15 |
| Psychology | 12 | 3 | 12 | 27 |
| Other | 10 | 11 | 10 | 31 |
| **Need for cognition scale (1 to 7 scale)** | | | | |
| Score | M = 3.79 (SD = 0.48) | M = 3.84 (SD = 0.62) | M = 3.65 (SD = 0.62) | M = 3.76 (SD = 0.58) |

Table 7: Articles, sources, and Figma prototypes used in the experiment

| Article, original source | Source link | Prototype link |
|---|---|---|
| Article 1, Russia Today | Source | Article 1 Basic Link |
| | | Article 1 Light Link |
| | | Article 1 Full Link |
| Article 2, Russia Today | Source | Article 2 Basic Link |
| | | Article 2 Light Link |
| | | Article 2 Full Link |
| Article 3, InfoWars | Source | Article 1 Basic Link |
| | | Article 3 Light Link |
| | | Article 3 Full Link |

Table 8: Net Promoter Score value breakdown for Light and Full groups

| NPS Value | Detractor | | | | | | | Passive | | Promoter | |
|---|---|---|---|---|---|---|---|---|---|---|---|
| | 0 | 1 | 2 | 3 | 4 | 5 | 6 | 7 | 8 | 9 | 10 |
| Light (N = 84) | 3 | 1 | 2 | 6 | 2 | 11 | 13 | 20 | 10 | 9 | 7 |
| Full (N = 83) | 2 | 0 | 2 | 4 | 1 | 7 | 5 | 10 | 21 | 11 | 20 |

Table 9: Experiment results for measures comparing three groups (Basic, Light, Full)

| Variable | Basic (N = 162) | | Light (N = 168) | | Full (N = 166) | | One-Way Anova | | Two-sided T-Test Basic vs Light | | | Two-sided T-Test Light vs Full | | | Two-sided T-Test Basic vs Full | | |
|---|---|---|---|---|---|---|---|---|---|---|---|---|---|---|---|---|---|
| | M | SD | M | SD | M | SD | F (2, 485) | p | t | df | p | t | df | p | t | df | p |
| **Reading time** | 150.705 | 180.063 | 137.707 | 94.366 | 199.4145 | 159.611 | 7.723 | .0005 | .825 | 328 | .409 | -4.279 | 324 | .000 | -2.558 | 318 | .011 |
| **Thinking Mode (Mean)** | 4.124 | 1.060 | 4.248 | .897 | 4.516 | 1.027 | 6.445 | 0.02 | -1.144 | 328 | .253 | -2.511 | 324 | .013 | .3.352 | 318 | <.001 |
| Speed | 4.784 | 1.614 | 4.929 | 1.466 | 5.235 | 1.599 | 3.584 | .0285 | .852 | 328 | .395 | -1.646 | 324 | .101 | -2.359 | 318 | .019 |
| Processing | 3.914 | 1.692 | 3.702 | 1.715 | 4.181 | 1.930 | 3.018 | .0498 | 1.126 | 328 | .261 | -2.196 | 324 | .029 | -1.145 | 318 | .253 |
| Control | 4.235 | 1.456 | 4.619 | 1.512 | 4.783 | 1.670 | 5.406 | .0048 | -2.352 | 328 | .019 | -.973 | 324 | .331 | -3.168 | 318 | .002 |
| Effort | 4.068 | 1.646 | 3.970 | 1.769 | 4.367 | 1.769 | 2.382 | .0934 | .519 | 328 | .604 | -2.074 | 324 | .039 | -1.621 | 318 | .106 |
| Nature | 3.562 | 1.520 | 3.762 | 1.552 | 3.952 | 1.665 | 2.496 | .0834 | -1.183 | 328 | .238 | -1.015 | 324 | .311 | -2.136 | 318 | .033 |
| Adaptability | 4.185 | 1.749 | 4.506 | 1.586 | 4.632 | 1.623 | 3.183 | .0423 | -1.75 | 328 | .082 | -.677 | 324 | .499 | -2.334 | 318 | .020 |
| **Perception of bias** | 4.611 | 1.608 | 5.143 | 1.556 | 4.982 | 1.574 | 4.895 | .0079 | -3.052 | 328 | .002 | .939 | 324 | .3482 | -2.110 | 318 | .035 |

Table 10: Experiment results for measures comparing two groups (Light, Full)

| Variable | Light(N = 84) | | Full(N = 83) | | One-WayAnova | |
|---|---|---|---|---|---|---|
| | M | SD | M | SD | F (1, 165) | p |
| Propaganda awareness | .82 | .385 | .96 | .188 | 9.185 | .003 |
| Net Promoter Score | 6.37 | 2.404 | 7.49 | 2.416 | 9.096 | .003 |

Table 11: Expert assessment

| Variable | Mean | Std. Deviation | Mean | Std. Deviation | Percentage agreement(adj. scale) |
|---|---|---|---|---|---|
| | 7-point Likert Scale | | Adjusted scale* | | |
| **Accuracy** | **4.71** | **1.275** | **2.36** | **0.613** | **44.9%** |
| Expert 1 | 4.74 | 1.662 | 2.35 | 0.847 | |
| Expert 2 | 4.68 | 2.008 | 2.37 | 0.906 | |
| **Informativeness** | **4.68** | **1.174** | **2.36** | **0.577** | **42.1%** |
| Expert 1 | 4.89 | 1.656 | 2.43 | 0.778 | |
| Expert 2 | 4.47 | 1.574 | 2.30 | 0.892 | |
| **Believability** | **5.03** | **1.124** | **2.51** | **0.547** | **52.3%** |
| Expert 1 | 4.81 | 1.672 | 2.40 | 0.799 | |
| Expert 2 | 5.24 | 1.504 | 2.63 | 0.707 | |
| **Clarity** | **5.06** | **1.126** | **2.50** | **0.523** | **47.7%** |
| Expert 1 | 5.23 | 1.708 | 2.52 | 0.769 | |
| Expert 2 | 4.89 | 1.513 | 2.48 | 0.793 | |
| | Binary (0 – 'no', 1 – 'yes') | | | | |
| **Technique identification** | **0.81** | **0.27** | - | - | **68.2%** |
| Expert 1 | 0.90 | 0.305 | - | - | |
| Expert 2 | 0.73 | 0.447 | - | - | |
| Total | 107 | | *Adjusted scale:* 1 – bad ($\leq 3$ on 7-point Likert scale), 2 – neutral, 3 – good ($\geq 5$ on 7-point Likert scale) | | |